\documentclass{frontiersSCNS} 

\usepackage{url,hyperref,lineno,microtype}
\usepackage[onehalfspacing]{setspace}

\newcommand{\ie}{{\em i.e.}}
\newcommand{\eg}{{\em e.g.}}
\newcommand{\sext}{\texttt{SExtractor}}
\newcommand{\apj}{{\em Ap.J.}}
\newcommand{\apjl}{{\em Ap.J. Lett.}}
\newcommand{\apjs}{{\em Ap.J. Suppl. Ser.}}
\newcommand{\aj}{{\em A.J.}}
\newcommand{\mnras}{{\em Mon. Not. R. Astr. Soc.}}
\newcommand{\aap}{{\em Astron. Astrophys.}}
\newcommand{\pasp}{{\em Pub. Astr. Soc. Pacific}}

\def\keyFont{\fontsize{8}{11}\helveticabold }
\def\firstAuthorLast{D'Onofrio {et~al.}} 
\def\Authors{Mauro D'Onofrio\,$^{1,*}$, Paola Marziani\,$^{2}$ and Lucio Buson\,$^2$}
\def\Address{$^{1}$Department of Physics and Astronomy, University of Padova, Padova, Italy \\
$^{2}$INAF, Astrophysical Observatory of Padova, Padova, Italy}
\def\corrAuthor{Mauro D'Onofrio}
\def\corrAddress{Department of Physics and Astronomy, University of Padova, Vicolo Osservatorio 3, 35122 Padova, Italy}
\def\corrEmail{mauro.donofrio@unipd.it}

\begin{document}
\onecolumn
\firstpage{1}

\title[The transformation of Spirals into S0 galaxies in the cluster environment]{The transformation of Spirals into S0 galaxies in the cluster environment} 

\author[\firstAuthorLast ]{\Authors} 
\address{} 
\correspondance{} 

\extraAuth{}

\maketitle

\begin{abstract}

\section{}

We discuss the observational evidences of the morphological transformation of Spirals into S0 galaxies
in the cluster environment exploiting two big databases of galaxy clusters: WINGS ($0.04<z<0.07$) and EDisCS ($0.4<z<0.8$).

The most important results are: 1) the average number of S0 galaxies in clusters is almost a factor of $\sim 3-4$ larger today than at redshift $z\sim 1$; 2) the fraction of S0's to Spirals increases on average by a factor $\sim 2$ every Gyr;
3) the average rate of transformation for Spirals (not considering the infall of new galaxies from the cosmic web) is: $\sim5$ Sp$\rightarrow$S0's per Gyr and $\sim2$ Sp$\rightarrow$E's per Gyr; 4) there are evidences that the interstellar gas of Spirals is stripped by an hot intergalactic medium; 5) there are also indirect hints that major/minor merging events have played a role in the transformation of Spiral galaxies. 

In particular, we show that: 1) the ratio between the number of S0's and Spirals ($N_{S0}/N_{Sp}$) in the WINGS clusters is correlated with their X-ray luminosity $L_X$; 2) that the brightest and massive S0's are always close to the cluster center; 3) that the mean S\'ersic index of S0's is always larger than that of Spirals (and lower than E's) for galaxy stellar masses above $10^{9.5} M_{\odot}$; 4) that the number of E's in clusters cannot be constant; 5) that the largest difference between the mean mass of S0's and E's with respect to Spirals is observed in clusters with low velocity dispersion. 

Finally, by comparing the properties of the various morphological types for galaxies in clusters and in the field, we find that the most significant effect of the environment is the stripping of the outer galaxy regions, resulting in a systematic difference in effective radius and S\'ersic index.

\tiny
 \keyFont{ \section{Keywords:} clusters of galaxies; galaxy formation; galaxy morphology} 
\end{abstract}

\section{Introduction}\label{Intro}

The morphological class of S0 galaxies has been largely debated since \cite{Hubble1936} supposed their existence in ``Realm of the Nebulae", and \cite{Sandage1961} placed them as transition types in the tuning fork diagram.
\cite{Spitzer&Baade1951} first suggested that S0's might originate from Spirals stripped out of gas and that this class of objects could form a dustless sequence that parallels the Sa-Sb-Sc sequence in the Hubble tuning fork.
This idea was further developed by \cite{vandenBergh1976} who proposed the class of ``anemic Spirals" as intermediate objects between normal S0's and Spirals.

Historically there are essentially two main views concerning the origin of S0 galaxies: 1) S0's are the end product of the transformation of Spiral galaxies removed of their gas content by means of gravitational or non gravitational forces, or by secular evolution; 2) the S0 class is intrinsic, \ie\ formed by objects that differ from Spirals since their formation.

The proponents of the first view found support in the following observational facts: 1)  a significant fraction of today's early-type galaxies likely evolved from later-type galaxies at relatively recent cosmic epochs.  Hubble Space Telescope (HST) observations revealed that Spirals are proportionally much more common (a factor of $\sim2-3$), and S0 galaxies much rarer, in distant than in nearby clusters 
\citep[see \eg][]{Dressleretal1997,Fasanoetal2000,Treuetal2003,Postmanetal2005,Smithetal2005,Desaietal07}; 
2) several evidences exist that Spirals are stripped of their HI gas in dense environments
\citep[see \eg][]{Davies&Lewis1973,Haynesetal1984,Giovanelli&Haynes1985,Solanesetal2001,Koopmann&Kenney2004,Crowletal2005,Chungetal2007,Sunetal2007,Yagietal2007,Rasmussenetal2008,Vollmeretal2009,Yoshidaetal2008,Smithetal2010,Sivanandametal2010}. Loosing their gas Spiral galaxies rapidly transform into S0's owing to the quenching of their star formation (SF)
\citep[see \eg][]{BoselliGavazzi2006,BookBenson2010,Fabelloetal2011,Goncalvesetal2012,Mendeletal2013,Bekki2014,Hirschmannetal2014,Wuetal2014}; 3) the specific frequency of globular clusters (GCs) in present day S0's is consistent with the idea that these galaxies are formed when the gas in normal Spirals is removed \citep{Barraetal2007}; 4) the observed mass-metallicity trends in S0 galaxies are consistent with a merging formation of these galaxies \citep{Chamberlainetal2012}; 6) the stellar kinematics of S0 galaxies inferred from Planetary Nebulae (PN) appears consistent with those of Spirals, indicating that they could evolve directly from such systems via gas stripping or secular evolution \citep{Cortesietal2011}.

On the other side the proponents of the second view claim that:
1) the bulge-to-disk ratios ($B/D$) of the two morphological classes are too different for being explained by the simple occurrence of gas stripping \citep[see \eg][]{Burstein1979}. If S0's are evolved Spirals, their bulge luminosities must have been physically enhanced \citep{Christlein&Zabludoff}; 2) many edge-on S0's are characterized by thick disks that are absent in the control samples of Spirals;  3) the relation between density and morphology is a weak
function of density, so that a significant percentage of S0's are in regions where the
density and temperature is too low for a significant stripping of the gas \citep{Dressler1980};
4) the large ages and magnesium-to-iron overabundance of the outer disks of nearby S0's contradict
the commonly accepted scenario of S0 formation from Spirals by quenching their
star formation during infall into dense environments at intermediate redshifts, $z = 0.4-0.5$
(4-5 Gyr ago) \citep{Silchenkoetal2012}; 5) the evolution in the number density of S0's from  $z = 0.4-0.5$
to  $z = 0$ is the result of morphological classification errors \citep{Andreon1998};
6) neither the median ellipticity, nor the shape of the ellipticity distribution of cluster early-type galaxies evolve with redshift from $z \sim 0$ to $z > 1$, \ie\ over the last $\sim 8$ Gyr \citep{Holdenetal2009}, suggesting that the bulge-to-disk ratio distribution for cluster early-type galaxies does not change with time. 

The controversy on the S0' origin is not yet settled after more than 70 years!
There is however today a wide consensus that more than one process can operate the transformation of Spirals into S0's and that a significant role is played by the environment.

In low density environments the transformation of Spirals into S0's could be the end product of secular evolution. This idea is supported by the fact that many S0's in this environment host pseudo-bulges \citep{Laurikainenetal2006,Graham2013}, \ie\ bulges with bulge-to-total luminosity ratios and concentrations typical of late-type spiral galaxies ($B/T < 0.2$ and $n \sim1$). These bulges often contain embedded disks, inner spiral patterns, nuclear bars, star formation, and 
present rotational support similar to that of Spiral bulges \citep[see][]{KormendyKennicutt2004}. These properties
are usually attributed to the evolution induced by bars, which send material towards the center of the galaxy, enhancing the bulge component \citep{PfennigerNorman1990}, even if often these bulges do not seem to have the concentrations and sizes of typical S0's bulges \citep[see \eg][]{Eliche-Moral2012,Eliche-Moral2013}.  However, 
\cite{PfennigerFriedli1993} found that the buckling instability can lead to vertically thick inner bar components, which are several factors larger than the nuclear star forming structures. They have
similar sizes as the boxy/peanut bulges in the S0s (and in other
Hubble types) in the edge-on view. Furthermore, secular evolution offers a viable mechanism to produce the thick disk component observed in many edge-on S0's.

In the cluster environment several mechanisms can operate the quenching of SF and start the transition of Spirals toward the S0 class: 1) gas loss triggered by: galactic winds as a consequence of star formation or AGN activity \cite[see, e.g.][]{Veilleuxetal2005,Hoetal2014,Fogartyetal2012}, ram pressure stripping, i.e. the interaction between the interstellar medium (ISM) of the galaxy and the intergalactic medium (IGM) \citep{Gunn&Gott1972}, strangulation, i.e.  the removal of the hot gas halo surrounding the galaxy via ram pressure or tidal stripping by the halo potential \citep{Larsonetal1980,Baloghetal2000}, thermal evaporation \citep{Cowie&Songaila1977}, and turbulent/viscous stripping \citep{Nulsen1982}. 2) strong tidal interactions and mergers, although more frequent in groups than in clusters \citep{Barnes&Hernquist1992,Mazzeietal2014a,Mazzeietal2014b}, tidal effects from the cluster as a whole \citep{Byrd&Valtonen1990}, and harassment, i.e. the cumulative effect of several weak and fast tidal encounters \citep{Mooreetal1996}.

Ram pressure is one of the most accredited actors of the transformation of Spirals in clusters. Its importance resides in the fact that only the gaseous component of disk galaxies is affected. The ram pressure exerted by the IGM is a key ingredient for understanding the star formation history (SFH) of galaxies in clusters \citep[see e.g.,][]{Bekki2014}, because temporal and spatial variations of the SF in disks can be produced by ram pressure \citep[see e.g.,][]{Koopmann&Kenney2004}. Observations suggest that the H$\alpha$-to-optical disk-size ratio is smaller for Spirals with higher degrees of HI-deficiency \citep{Fossatietal2013}. 

The merging hypothesis is also quite popular and old (see e.g., \cite{BiermannTinsley}; mechanism defined at that time "collision between galaxies") and often advocated for explaining the existence of S0 galaxies in low density environments (e.g. groups and field) where the mechanisms typical of clusters are less effective. This possibility found support in the discovery of "Polar Ring S0's", i.e. lenticular galaxies hosting structures likely due to a second event such as the transfer of mass from a companion galaxy during a close encounter as well as a true merger \citep{SchweizerWhitmoreRubin1983}. \cite{Bekki1997} stressed that a particular orbit configuration could transform two late-type spirals into one S0 galaxy with polar rings \citep[see also][]{MapelliRampazzoMarino2015}. A related phenomenology, namely S0's hosting an unexpected gas component counter-rotating or strongly kinematically decoupled with respect to the stellar body of the galaxy, has also been detected \citep{BertolaBusonZeilinger1992}. Even more, numerical simulations by \cite{Bekki1998} indicate that the major mechanism for the S0 creation is galaxy merging between two spirals with unequal mass; if this is true, unequal-mass galaxy mergers provide the evolutionary link between the large number of blue spirals observed in intermediate-redshift clusters and the red S0's in the present-day ones. 

An additional piece of evidence favoring the merging idea comes from the correlation observed in very nearby galaxies between UBV color residuals and the Schweizer's fine-structure $\Sigma$ index, especially because the detected systematic variations are not limited to the nuclei, but occur globally in the stellar populations of galaxies \citep{SchweizerSeitzer1992}. Again with reference to observed colors, the so-called blue E/S0 galaxies studied by \cite{AguerriHuertas-CompanyTresse2010}, i.e. objects having a clear early-type morphology on HST/ACS images but with a blue rest-frame color, do resemble merger remnants probably migrating to the red-sequence in a short time-scale, when their mass is comparable (or higher) than that of the Milky Way. Analogously \cite{Prietoetal2013} identify several pieces of evidence favoring the same hypothesis i.e. that major mergers have played a dominant role in the definitive buildup of present-day E-S0's with $M* > 10^{11} M_{\odot}$ at $0.6 < z < 1.2$. 

At any rate the frequency of other photometric hints indicating a major role of merging events in S0 galaxies formation is still controversial. For instance the claim by \cite{Borlaffetal2014}, who find that---in the context of their N-body simulations---nearly 70\% of their S0-like host anti-truncated (Type~III) stellar disks does not find support in the observational works by \cite{Erwinetal2008} as well as by
\cite{Gutierrezetal2011}, who at variance conclude that the difference in the fractions of Type-III profiles between early (presumably S0-S0a) 
and late-type spirals is not statistically significant. in other words, it should not reasonable to conclude that Type-III photometric profiles are a special characteristic of S0's. \cite{Maltbyetal15} reports for a sample
of $\sim280$ field and cluster S0's similar fractions of Type profiles in both environments.


On the other side \cite{Querejetaetal2015} have shown that major mergers among binary galaxies can rebuild a bulge-disk coupling in the S0 remnants in less than $\sim$3 Gyr, after having destroyed the structures of the progenitors, whereas minor mergers directly preserve them. The resulting structural parameters span a wide range of values, and their distribution is consistent with observations. In their simulations many remnants have bulge S\'ersic indices ranging $1 < n < 2$, flat appearance,  and contain residual star formation in embedded discs, a result which agrees with the presence of pseudo-bulges in real S0's. 
\cite{Querejetaetal2015b} also find that major mergers can explain the difference in the stellar angular momentum and concentration between Spirals and S0's.

Disentangling the mechanisms at work in the transformation of Spirals into S0 galaxies is not easy even in very nearby objects. For more distant galaxies a combination of efforts are required, both in the sense of increasing the observational data available for galaxies in clusters at different redshifts, and in producing numerical simulations of increasing complexity. 
With this paper we want to contribute to this effort, discussing the data coming from two clusters surveys, WINGS \citep[WIde field Nearby Galaxy cluster Survey;][]{Fasanoetal2006} at redshifts $0.04\leq z\leq 0.07$ and EDisCS \citep[the ESO Distant Cluster Survey;][]{Whiteetal05} at $0.4\leq z\leq 0.8$, that are particularly important for their ample databases. The idea behind this work is that by comparing the mean properties of galaxies of different morphological classes per bin of masses, cluster-centric distance, and redshift, we can find a statistically significant proof for the origin of the S0 class.

We start with Sec.~\ref{Sec1} and its subsections discussing the WINGS and EDisCS data samples used for the present analysis, as well as the characteristics of three additional subsamples that have been used to support our conclusions. In Sec.~\ref{Sec2} we provide the main observational evidences in favor of the hypothesis that Spirals in clusters transform into S0's. In Sec.~\ref{Sec3} we present a detailed analysis of the stacked WINGS and EDisCS data samples discussing their implication for the origin of the S0 class. In Sec.~\ref{Sec4} we discuss the evidences of ram pressure stripping and evaporation, and in Sec. ~\ref{Sec5} those of merging. In Sec.~\ref{Sec6} we compare the properties of field and cluster S0's and,
finally in Sec.~\ref{Sec7} we present  our conclusions.

Throughout the paper we have adopted the following cosmology: $H_0=70\, km s^{-1} Mpc^{-1}$, $\Omega_M =0.3$ and $\Omega_{\Lambda}=0.7$.


\section{The data samples of galaxies in clusters}\label{Sec1}
\subsection{WINGS}\label{Sec1a}

WINGS is a long-term project dedicated to the characterization of the photometric and spectroscopic properties of galaxies in nearby clusters.  The core of the survey is WINGS-OPT \citep{Varelaetal2009}, which is a set of B and
V band images of a complete, X-ray selected sample of 77 clusters with
redshift $0.04<z<0.07$.  The images have been taken with the Wide
Field Camera (WFC, $34'\times 34'$) at the INT-2.5 m telescope in La
Palma (Canary Islands, Spain) and with the Wide Field Imager (WFI,
$34'\times 33'$) at the MPG/ESO 2.2 m telescope in La Silla
(Chile). The optical photometric catalogs have been obtained using
\sext\ and are 90\% complete at V$\sim$21.7, which translates into
$M^{*}_{V} + 6$ at the mean redshift of the survey \citep{Varelaetal2009}.
The WINGS-OPT catalogs contain $\sim$400,000 galaxies in both the V-
and B-band. According to \citet[][Table~D.2]{Varelaetal2009}, in the
whole cluster sample the surface brightness limits at
1.5$\sigma_{bkg}$ ($\sigma_{bkg}$ is the standard deviation per
pixel of the background) span the ranges $24.7\div26.1$ (average value:
25.71) in the V band  and $25.4\div26.9$ (average value: 26.39) in the B band.
\sext\ catalogs have also been obtained for the near-infrared
follow-up of the survey \citep[WINGS-NIR;][]{Valentinuzzietal2009}, which consists
of J- and K-band imaging of a subsample of 28 clusters of the
WINGS-OPT sample, taken with the WFCAM camera at the UKIRT
telescope. Each mosaic image covers $\approx$0.79$\rm{deg}^2$. With
the \sext\ analysis the 90\% detection rate limit for galaxies is
reached at J=20.5 and K=19.4. The WINGS-NIR catalogs contain
$\sim$490,000 in the K band and $\sim$260,000 galaxies in the J band.
The photometric depth of the WINGS-NIR imaging
is slightly worse than that of the WINGS-OPT
imaging. Thus, for the K-band the surface brightness limit at
1.5$\sigma_{bkg}$ spans the range $20.6\div21.5$ \citep[see Table 4 in][average value: 21.15]{Valentinuzzietal2009}.
The U-band photometry  has been realized up to now for a subsample (17 clusters) of galaxies of the WINGS-OPT
sample \citep[WINGS-UV;][]{Omizzoloetal2014}, but new $u$ band imaging of the WINGS clusters have been recently acquired at the VST telescope
within  OMEGAWINGS project, that is an extention of WINGS.
The area covered is $\sim1$ square degree and allows to reach up to 2.5 cluster virial radii. OMEGAWINGS is based
on two OmegaCAM@VST GTO programs for 46 WINGS clusters. The B and V campaigns of observations are already completed, while the
$u$-band programme is still ongoing. The B and V data of OMEGAWINGS are presented in Gullieuszik et
al. (2015). 

In addition to the aperture photometry catalogs, the WINGS database includes the surface photometry catalogs realized with GASPHOT \citep{Donofrioetal2014}.
This automatic software provided for each galaxy and each band (with an area larger than 200 pixels) the total luminosity,
the effective radius $R_e$, the effective surface brightness $\langle \mu \rangle_e$, the Sersic index $n$,
and the minor to major axis ratio $b/a$.
In addition the automatic tool MORPHOT \citep{Fasanoetal2012}
produced the accurate morphological classification $T$ of $\sim$40,000 galaxies.
For more information about the WINGS database see \cite{Morettietal2014}.

The medium-resolution multifiber spectroscopy of 48 clusters is another valuable property of the WINGS database \citep[WINGS-SPE;][]{Cavaetal2009}.
Redshifts and membership were measured \citep[see for details,][]{Cavaetal2009}, 
as well as star formation histories, stellar masses, ages \citep[see for details,][]{Fritzetal2007,Fritzetal2011}, equivalent widths, and line-indices \citep{Fritzetal2014} for $\sim$6,000 galaxies.
Within OMEGAWINGS new spectra have been obtained with AAOmega@AAT using the OmegaCAM fields. So far, we have secured high quality
spectra for $\sim30$ OMEGAWINGS clusters, reaching very
high spectroscopic completeness levels for galaxies brighter
than $V=20$ from the cluster cores to their periphery (Moretti
et al. in prep.).

\subsection{The WINGS stacked sample}\label{Sec1b}

Using all the galaxies members of the 48 WINGS clusters with available spectroscopy\footnote{The cluster membership was measured on the basis of the velocity dispersion of the galaxies in the clusters \citep[see,][]{Cavaetal2009}.} we have built a stacked sample aiming at
increasing the statistical significance of the results obtained  for the different morphological classes. 

We extracted from the WINGS database the following data: the identification name (ID), the cluster name, the redshift  ($z$), the scale of the images (kpc/arcsec),  the value of $R_{200}$\footnote{$R_{200}$ is the radius delimiting a sphere with interior mean density 200 times the critical density of the Universe at that redshift \citep[see for details][]{Cavaetal2009}} in Mpc obtained for the cluster, the morphological type ($T$), the luminosity distance ($D_{lum}$), the equatorial coordinates of the galaxies ($RA$ and $DEC$), the B and V GASPHOT total magnitudes, the circularized effective radius ($R^c_e$) in kpc unit, the mean effective surface brightness ($\langle \mu \rangle_e$) in mag arcsec$^{-2}$, the distance from the brightest cluster member  ($D_{BCG}$) in $R_{200}$ unit, the star formation rate in the last $2\times10^7$ Myr ($SFR[M_{\odot}yr^{-1}$]), the axis ratio in the V band ($b/a$), the total stellar mass ($M_* [\odot]$), the luminosity weighted age ($wAge[yr]$), the local density ($LD$), the galactic extinction in the B and V bands, the S\'ersic index ($n$) in the V band, and the average B$-$V color.

In total we got 2869 galaxies members of the clusters from a total number of objects of the spectroscopic catalog of $\sim 6000$ galaxies.
According to their morphological classification we have 803 Elliptical galaxies
($T<-5$), 1347 S0 galaxies ($-4.5<T<0$), and 719 ($T>0$) Spiral galaxies. They span an interval of masses from $\sim 10^8$ to $\sim 10^{12}$ $M_{\odot}$, and an interval of distances from the center from 0 to $\sim1$ $R_{200}$.

The stacked WINGS sample is aimed at creating a ``pseudo-cluster'' having the mean properties of our clusters, but with many more objects at each distance $D_{BCG}/R_{200}$ from the center. This is possible because the WINGS clusters share well defined properties in terms e.g. of X-ray luminosities, velocity dispersions, dimensions, richness, ages, etc. This homogeneity of properties is also accompanied by a set of homogeneous observations and data reductions, so that the final sample is very robust from a statistical point of view. We can say that this pseudo-cluster
provides us the mean properties of the WINGS clusters and its variance. This stacked sample permits the analysis of the mean properties of the morphological classes as a function of mass and distance from the cluster center.

The completeness of this data sample is the same of our spectroscopic sample. This is defined by the ratio of the number of spectra yielding the redshift to the total number of galaxies in the photometric catalogue. \cite{Cavaetal2009} found that this sample 
is $\sim$50\% complete at V$\sim18$ mag. This corresponds to $M_V \sim -19$ mag and $\log(M_*/M_{\odot}) \sim10$. The spectroscopic completeness rapidly decreases below V$\sim19$ mag. 
The completeness turns out to be rather flat with magnitude for most clusters, and it is essentially independent from the distance to the cluster center up to $\sim0.6 R_{200}$. 

\begin{figure}
\center
\includegraphics[angle=0,scale=0.8]{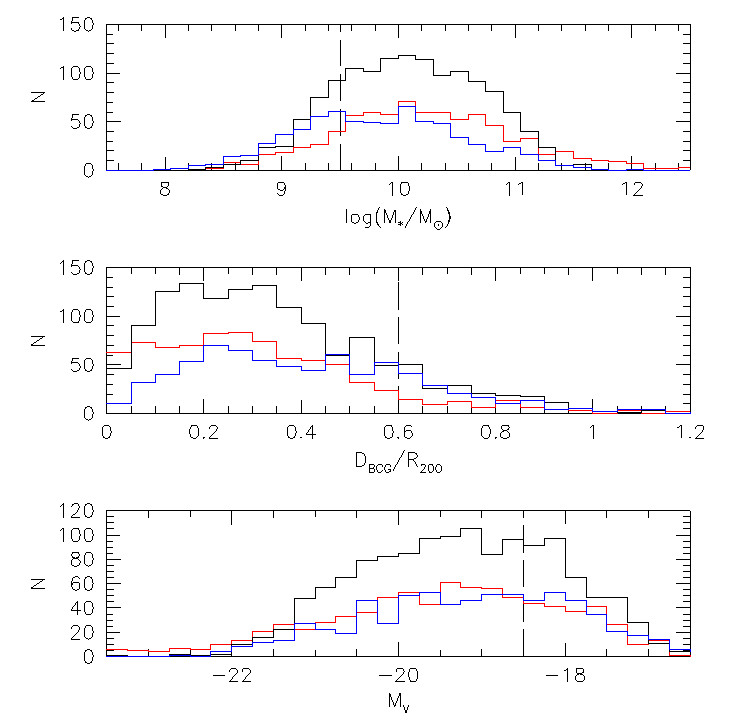}
\caption{From top to bottom  we plot here the distribution of the galaxy stellar masses,  the cluster-centric distance in $R_{200}$ units and the absolute magnitudes of the galaxies for the different morphological types: 
red lines mark the Elliptical galaxies, black lines the S0 galaxies, and blue lines Spirals. The vertical dashed black lines mark the limit where the WINGS data sample is $\sim50\%$ complete.}
\label{fig1}
\end{figure}

\begin{figure}
\center
\includegraphics[angle=0,scale=0.8]{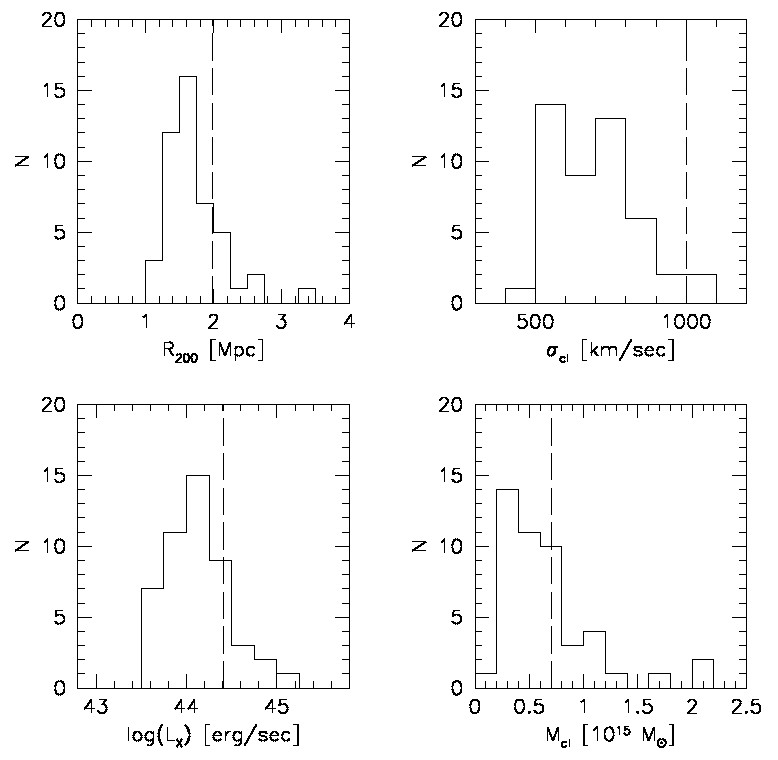}
\caption{From top left to bottom right we plot here the observed distributions of $R_{200}$, $\sigma_{cl}$, $\log(L_X)$ and $M_{cl}$ for the WINGS clusters that belong to the stacked sample. The dashed line mark the values for the Coma cluster.}
\label{fig2}
\end{figure}

Fig.~\ref{fig1} shows the distribution of the galaxy stellar masses, of the cluster-centric distance from the BCG normalized to $R_{200}$, and of the absolute magnitudes for the different morphological types. Note that the histograms of the masses and luminosities remain flat up to
$\log(M_*/M_{\odot}) \sim9.5$ and $M_V\sim-18$ mag respectively.
There are also no reasons to believe that the morphological fraction (\ie\ the ratios among E-S0-Sp types) drastically change when  the completeness decreases, and that the survey has systematically lost galaxies of a given morphological type in the explored interval of cluster-centric distances $0\leq D_{BCG}/R_{200}\leq 0.6$. We are therefore confident that up to these values of masses and luminosities the WINGS stacked sample is reasonably statistically robust.

Fig.~\ref{fig2} plots the distributions of $R_{200}$, $\sigma_{cl}$ (the central velocity dispersion of the cluster), $\log(L_X)$ (the X-ray luminosity of the cluster) and $M_{cl}$ (the dynamical mass of the clusters) for the WINGS clusters that belong to the stacked sample. The values for the Coma clusters are indicated by the dashed lines as a reference. These data are useful to better figure out the observed range of properties of the stacked cluster sample. Clusters of quite different properties have been averaged together.

\subsection{Additional subsamples used in this work}\label{sec1c}

Three smaller subsamples have been used in this work for testing some of our hypotheses. The first one contains the measured Lick indexes for 2241 WINGS galaxies in common with our stacked sample. The Lick indexes were measured by Alexander Hansson in his PhD thesis following the prescriptions of \cite{Worthey94}. These data are still unpublished. The second subsample contains the bulge-to-disk decomposition of thousand of galaxies (but only 877 in common with our stacked sample) measured by Ruben Sanchez-Janssen in his PhD thesis. All galaxies were 2D decomposed with a S\'ersic bulge and an exponential disk component. These data are also still unpublished.
The third subsample include 1579 galaxies with redshift $z\leq 0.1$ extracted from the B-band PM2GC survey \citep[see][for more details]{Calvietal2011,Calvietal2012}. These galaxies belong to low density environments and have been used to compare the properties of the different morphological classes in clusters and in the field. These galaxies have been analyzed adopting the same softwares (e.g. GASPHOT and MORPHOT) and procedures used in the data reduction of the WINGS data, so that there is a high level of homogeneity between the two databases. Clearly, all these subsamples are not complete from a statistical point of view, in particular for faint galaxies, but still they provide significant indications of the mean trends of the measured structural parameters.
We will explicitly mention the use of these subsamples when they are used.

\subsection{The high-z EDisCS stacked sample}\label{Sec1d}

The properties of the EDisCS data survey \citep{Whiteetal05} have been extensively analyzed in various papers \citep[see e.g.,][]{Vulcanietal2011b}.
This multiwavelength photometric and spectroscopic survey includes several clusters at $0.4<z<1$. The galaxy photometric redshifts were determined 
by using both optical and IR imaging \citep{Pelloetal09}, the morphologies were derived by \cite{Desaietal07}, and  the photometric parameters (radii, total magnitudes, colors, etc.) were extracted using \sext\ by \cite{Whiteetal05}.

For the comparison with WINGS we used the  mass-limited sample at $M_* = 10^{10.2} M_{\odot}$ derived  by \cite{Vulcanietal2011b}, consisting of 489 galaxies, 156 of which are classified as Es, 64 as S0s, 252 as Spiral galaxies and 17 with unknown morphology. The values of $R_{200}$ and the distance of the galaxies from the BCG were derived by \cite{Poggiantietal06,Poggiantietal08}.
The values of the galaxy stellar masses (still unpublished) were kindly provided by B. Vulcani: they are calculated using the Kroupa \citep{Kroupa01} IMF in the mass range $0.1-100 M_{\odot}$ following \citet{BelldeJong2001}. 

As for WINGS, the EDisCS data were also stacked together at increasing distance from the BCGs with the aim of providing
a more robust statistical comparison between high-z and low-z clusters.  

\section{Observational elements in favor of the transformation of Spirals into S0's}\label{Sec2}

Here we present the main observational evidences in favor of the hypothesis that Spirals transform themselves into S0's in clusters.

\begin{figure}
\includegraphics[angle=0,scale=0.45]{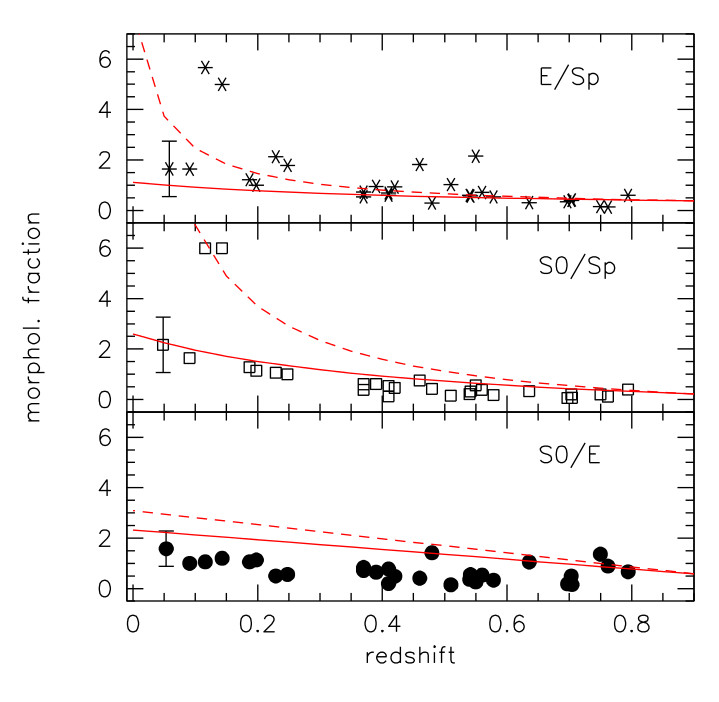}
\includegraphics[angle=0,scale=0.45]{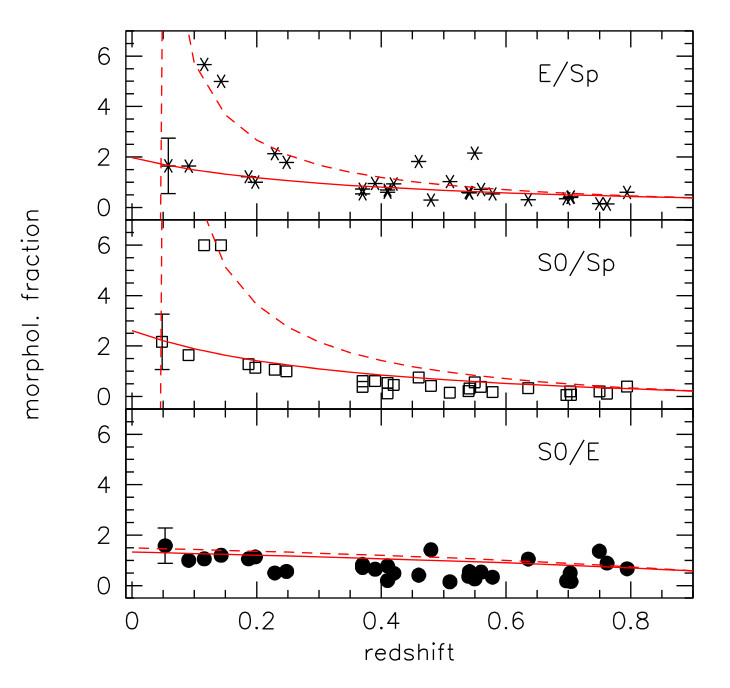}
\caption{Trend of the morphological fractions (E/Sp, S0/Sp and S0/E) with redshift in various clusters.  The WINGS stacked sample is shown with its $1\sigma$ error bar.
The plotted lines mark the evolution of the morphological fractions for a hypothetical cluster that at high redshift
has the mean morphological fractions of the EDisCS sample. 
Spirals transform into S0's and E's at different rates per interval of redshift. {\it Left panel:} The solid lines mark the
trend obtained for a rate of transformation of 6.25 Sp$\rightarrow$S0's per Gyr and 0 Sp$\rightarrow$E's per Gyr. The dashed lines instead are for a rate of 9 Sp$\rightarrow$S0's per Gyr and 0 Sp$\rightarrow$E's per Gyr. {\it Right panel:} The same plot, but with the following transformation rates: 5 Sp$\rightarrow$S0's per Gyr and 1.75 Sp$\rightarrow$E's per Gyr for the solid lines, and 7 Sp$\rightarrow$S0's per Gyr and 1 Sp$\rightarrow$E's per Gyr for the dashed lines.  
}
\label{fig3}
\end{figure}

Fig.~\ref{fig3} shows the fraction of S0/E, S0/Sp and E/Sp for all the clusters observed up to $z\sim 0.8$, coming from the
WINGS and EDisCS databases, as well as from literature data for the intermediate redshift ($0.1< z <0.55$)
clusters studied by \cite{Fasanoetal2000} and \cite{Dressleretal1997}. 

In their comparative analysis of the morphological fractions of clusters at low and high redshift \cite{Fasanoetal2000} already claimed the existence of a general trend in the morphological fractions (\%E's, \%S0's, \%Sp's) and in the S0/E and S0/Sp ratios with redshift, clearly suggesting that the S0 population tends to grow at the expense of Spirals, whereas the frequency of E's remains almost constant.

In Fig.~\ref{fig3} the WINGS clusters are represented by one single point marking the average value found for our stacked clusters sample and its variance. There are several things that are worth noting in this figure: 1) the S0/E ratio (bottom panel) is almost constant with redshift with
a significant variance; 2) the variance of the WINGS stacked sample is of the same order of the spread observed among clusters at all redshifts in all panels; 3) the S0/Sp ratio (mid panel) slightly increases at lower redshifts ($z<0.3$). This fraction follows the S0/E ratio but with a smaller dispersion up to $z\sim0.15$, where a strong peak is observed for two clusters A389 and A951; 4) the E/Sp fraction follows the same trend of the S0/Sp ratio, but with a bigger dispersion at all redshifts. Again a peak is observed at $z\sim0.15$ for the same clusters; 5) the S0 population dominate at $z\leq0.1$. Depending on the cluster we can have a factor of $\sim 3-4$ more S0's than Spirals;  6)
at redshift around 0.1 and below the number of Spirals observed is very small. 

The spread observed in the morphological fractions can be explained remembering that: a) clusters
have different structures and are more or less virialized/concentrated \citep{Oemler1974}; b) Spirals galaxies, that are the dominant population in the field, continuously enter into clusters from the cosmic-web filaments; c) there are possible errors in the classification of galaxies. 

\begin{figure}
\center
\includegraphics[angle=0,scale=0.6]{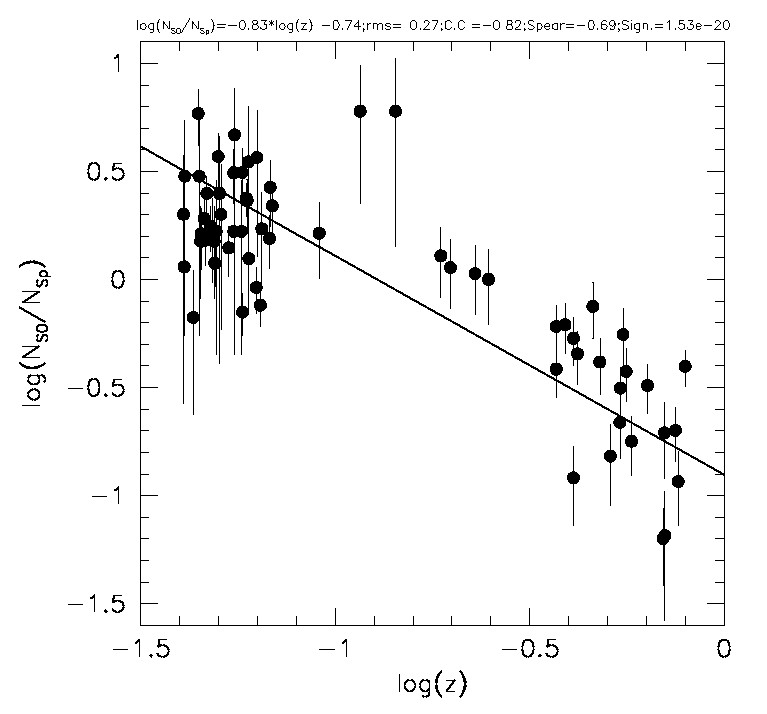}
\caption{The log of the $N_{S0}/N_{Sp}$ ratio versus the log of the redshift $z$. The solid line gives the best fit. Poissonian error bars mark the $1\sigma$ uncertainty of the observed counts. 
}
\label{fig4}
\end{figure}

The general impression coming from these data is that the effects of the morphological transformation become visible at redshift lower than $\sim0.3$, where a significant increase of the E/Sp and S0/Sp ratios are observed.
Despite the spread, the data seem to suggest an evolution with redshift of the morphological fractions.
If we trust in these data (i.e. that no systematic biases are present, e.g. in the morphological classification and in the surveyed area\footnote{The cluster area observed is around 1 Mpc$^2$ in almost all clusters.}), 
we must conclude from Fig.~\ref{fig3} that there is an active mechanism transforming Spirals into S0 objects (and possibly into E's). In fact, the ratios S0/Sp and E/Sp seem to increase with approximately the same gradients, although with quite different dispersion. 

The lines shown in Fig.~\ref{fig3} follow the hypothetical evolution of a cluster containing 26 E's, 15 S0's and 68 Spirals at redshift 0.9 (these values represent the average observed frequencies of morphological types in EDisCS). They mark the results of a simple calculation in which Spirals are transformed instantaneously into S0's and E's with different rates per interval of redshift. In the left panel the solid lines mark the transformation rates of 6.25 Sp$\rightarrow$S0 per Gyr and 0 Sp$\rightarrow$E per Gyr, while the dashed lines mark the rates of 9 Sp$\rightarrow$S0 per Gyr and 0 Sp$\rightarrow$E per Gyr. In the right panel  we have instead for the solid lines 5.0 Sp$\rightarrow$S0 per Gyr and 1.75 Sp$\rightarrow$E per Gyr, while for the dashed lines 7 Sp$\rightarrow$S0 per Gyr and 2.5 Sp$\rightarrow$E per Gyr. These rates were chosen to fit the general trends observed and to demonstrate that we can in principle follow the evolution of the morphological fraction of any cluster across time.
The observed trends demonstrate that an evolution of the morphological classes is necessary to explain the
data. Our calculation however does not take into account the infall, \ie\ the fact that new Spirals
enter the clusters across the cosmic time. Therefore, the rates calculated here should be considered lower limits.

Note that if we do not consider the transformation of Spirals into E's, we could not reproduce the mean trend of the S0/E fraction: both the low (solid line) and high (dashed line) rates of Spiral transformation into S0's chosen in the left panel, in which the number of E's remains constant, are not able
to fit the mean WINGS point giving the S0/E fraction.
On the other hand, when the transformation into E's is properly taken into account (see the right panel), we can reproduce both the mean trend of all morphological fractions and the peaks in the E/Sp and S0/Sp ratios\footnote{These peaks are not difficult to explain with a slightly higher frequency of transformation that can occur in some clusters in which the mechanisms of transformation are particularly efficient.} as well as the nearly constant S0/E ratio observed for all clusters.
In other word, or new E's enter systematically in the clusters or new E's must be formed through major merging of Spirals.

Fig.~\ref{fig4} shows the log of the $N_{S0}/N_{Sp}$ ratio versus the log of the redshift $z$. The least square fit is highly significant and demonstrate that the actual data can be interpreted in term of a temporal evolution of the morphological fraction, in the sense that S0 galaxies continuously increase in number with respect to Spirals. Transforming $z$ in age of the Universe with a linear approximation in the redshift range $0-1$, we see that the fraction $N_{S0}/N_{Sp}$ increases by a factor of $\sim 2$ every Gyr.

In conclusion, if our interpretation of Fig.~\ref{fig3} and ~\ref{fig4} are correct, Spirals transform into S0's and into E's with a slightly different rate. \cite{Vulcanietal2011b}, working on the same data for both surveys, arrived to the same conclusion with their analysis of the morphological fractions and mass functions.

\begin{figure}
\includegraphics[angle=0,scale=0.8]{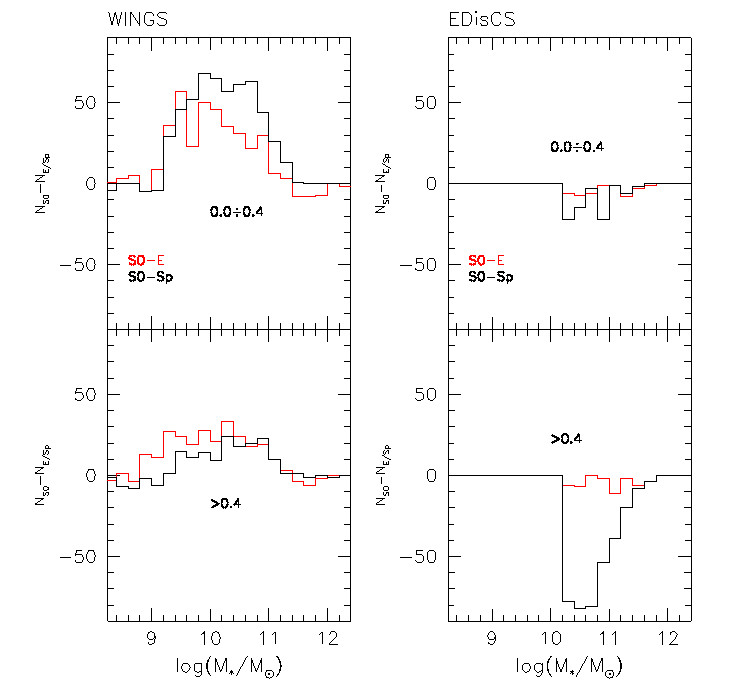}
\caption{The left panels plot with a black line the difference between the number of S0's and Spirals in the WINGS data sample for $D_{BCG}/R_{200}<0.4$ (upper panel)
and  $D_{BCG}/R_{200}>0.4$ lower panel in each bin of galaxy mass. The right panels show the same difference for the EDisCS data sample. The red lines mark the same distributions for the difference between the number of S0's and E's.}
\label{fig5}
\end{figure}

Fig.~\ref{fig5} provides us another hint in favor of the morphological evolution. It shows that S0's and Spirals are not spatially distributed in the same way in clusters at low and high redshifts. The figure clearly demonstrates that in the WINGS clusters S0's dominate
the central region of the clusters ($D_{BCG}/R_{200} < 0.4$), while Spirals in EDisCS are more frequent and abundant in the outer regions $D_{BCG}/R_{200} > 0.4$.
This fact, which reminds the morphology-density relation, is also consistent with the idea that Spirals transform into
S0's during their progressive infall toward the cluster central region.

\section{Dissecting the properties of the stacked cluster samples}\label{Sec3}

In this section we analyze in detail the properties of the different morphological types as a function of mass and
distance from the cluster center. The idea is that we can deduce from the observed differences of present day objects the mean path of the morphological transformation. In other words we can in principle establish which is the mean variation between two morphological classes.

\begin{figure}
\includegraphics[angle=-90,scale=0.6]{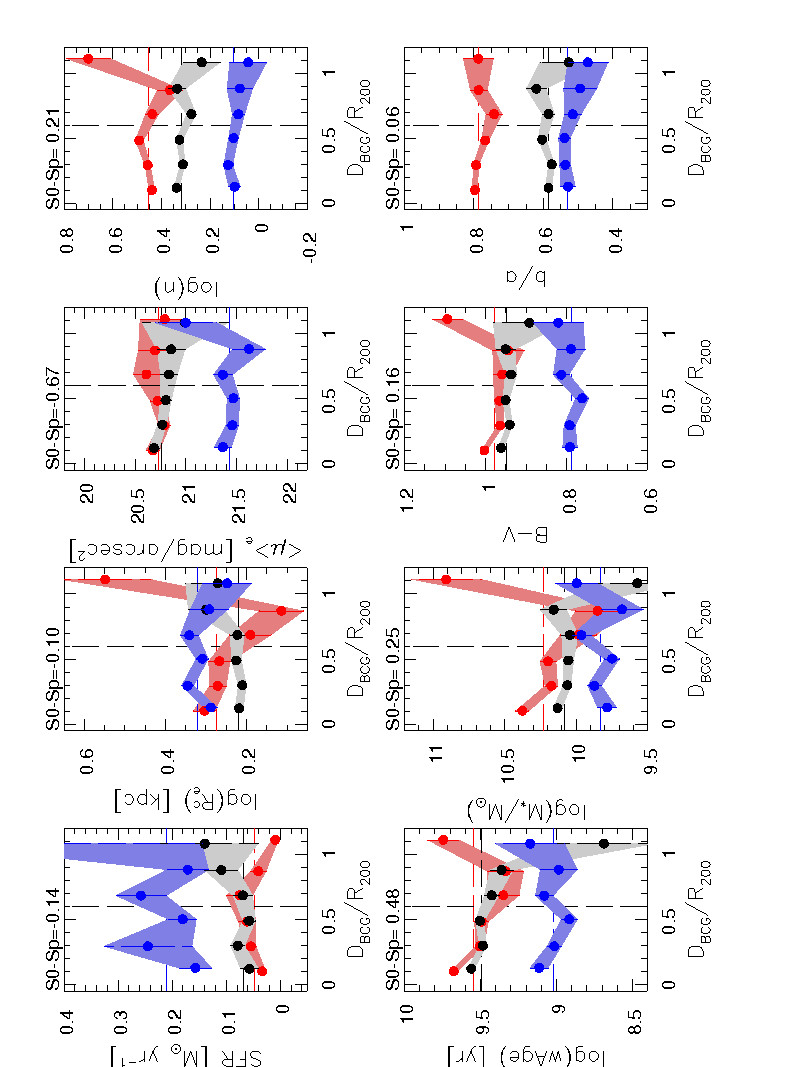}
\caption{From top left to bottom right we plot here versus the cluster-centric distance in $R_{200}$ units the following average quantities derived
for the different morphological types in various bins of $D_{BCG}/R_{200}$: the star formation rate (SFR), the log of the circularized effective radius
($\log(R^c_e[kpc])$, the mean effective surface brightness ($\langle \mu \rangle_e$), the log Sersic index ($\log(n))$, the log of the luminosity weighted age ($\log(wAge)$), the log total mass ($\log(M_{\odot}$), the B$-$V color, and the axis ratio ($b/a$). Red dots mark the Elliptical galaxies, black dots the S0 galaxies, and blue dots Spirals. The horizontal dashed lines provide the average value of the morphological classes, while the colored strips mark the
width of the error of the mean. The vertical dashed lines mark the limits where the stacked sample is $\sim50\%$ complete. On the top of each diagram we report the mean difference between the S0 and Spiral population.}
\label{fig6}
\end{figure}

Fig.~\ref{fig6} shows the results of our analysis for the WINGS stacked cluster sample.  From top left to bottom right we have plotted here the following quantities derived for the different morphological types $T$ in various bins of $D_{BCG}/R_{200}$: the star formation rate (SFR), the log of the circularized effective radius ($\log(R^c_e[kpc]))$, the mean effective surface brightness ($\langle \mu \rangle_e [mag\, arcsec^{-2}]$), the log S\'ersic index ($\log(n))$, the log of the luminosity weighted age ($\log(wAge[yr])$), the log total mass ($\log(M_\ast/M_{\odot}$)), the B$-$V color, and the axis ratio ($b/a$). Red, black and blue circles mark  the mean value of each quantity in each bin of distance respectively for E's, S0's and Spirals. We decided to plot with colored strips the error of the mean $\sigma/\sqrt{N}$, where $\sigma$ is the standard deviation in the bin and $N$ is the number of points, instead of the semi-interquartile in order to better highlight the mean differences among the various morphological classes. The width of the semi-interquartile is in fact $\sim 3$ times larger, so that in some cases the three morphological types share exactly the same parameter. In each diagram E's, S0's and Spirals are largely superposed, but the mean value for each population is clearly distinguishable. 

Apart from the clear separation between S0's and Spirals in each plot, and the general similarity of S0's with E's rather than with Spirals (with the notable exception of the S\'ersic index $n$), there are two things that are worth noting in this figure. First, S0's masses are on average almost a factor of $\sim 2$ larger than Spiral masses up to $\sim 0.6R_{200}$; second, S0's are $\sim 25\%$ smaller in their effective radius than Spirals within the same interval of cluster-centric distance.

\begin{figure}
\includegraphics[angle=-90,scale=0.6]{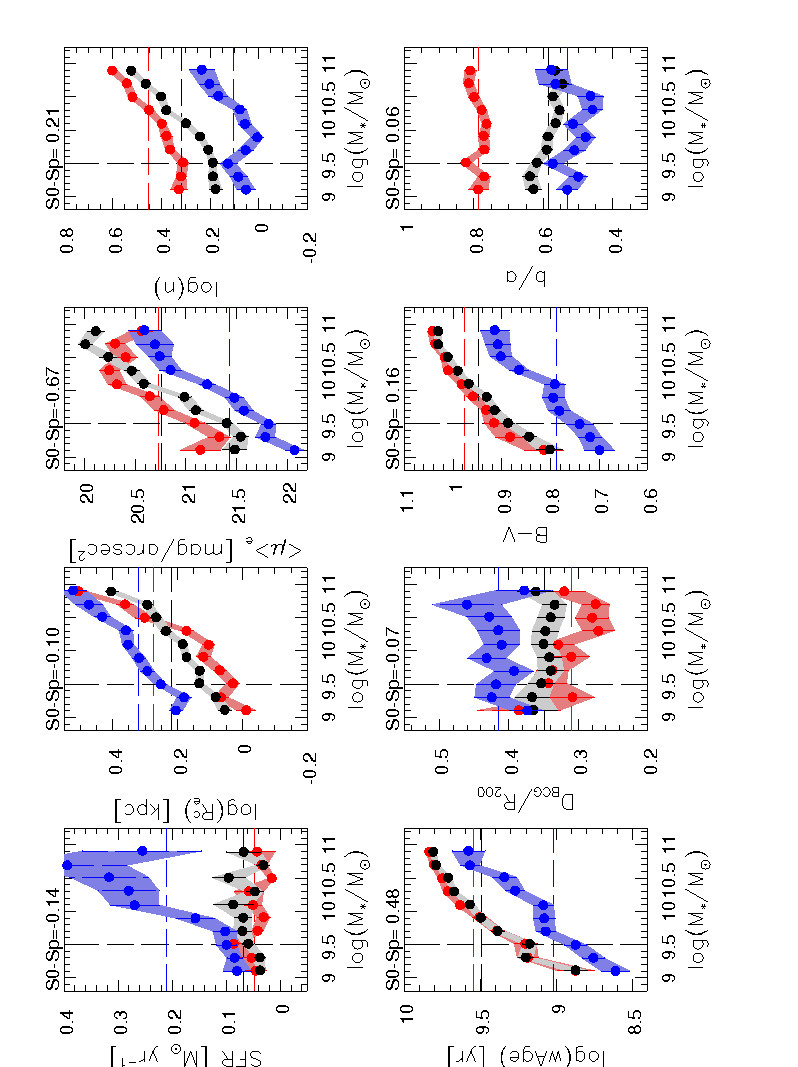}
\caption{From top left to bottom right we plot here versus the log mass of the galaxies in solar units the following average quantities derived
for the different morphological types in various bins of $D_{BCG}/R_{200}$: the star formation rate (SFR), the log of the circularized effective radius
($\log(R^c_e[kpc])$, the mean effective surface brightness in $R_{200}$ units ($\langle \mu \rangle_e$), the log Sersic index ($\log(n))$, the log of the luminosity weighted age ($\log(wAge)$), the distance from the BCG in  ($D_{BCG}/R_{200}$), the B$-$V color, and the axis ratio ($b/a$). Color codes as in Fig.~\ref{fig1}. 
The horizontal dashed lines provide the average value of the morphological classes, while the colored strips mark the
width of the error of the mean.
The vertical dashed lines mark the limits where the stacked sample is $\sim50\%$ complete. 
On the top of each diagram we report the mean difference between the S0 and Spiral population.}
\label{fig7}
\end{figure}

Fig.~\ref{fig7} shows a different view of the same WINGS data. This time we averaged the same quantities per bin of galaxy mass. The same color code marks again the three families of morphological types. Note that again S0's are very similar to E's in each plot, and that at every galaxy mass
E's are systematically closer to the cluster center than S0's and Spirals. This is a manifestation of the morphology-density relation \citep{Dressler1980}. Furthermore, S0's are at each mass on average $\sim 25\%$ smaller than Spirals, as seen before. Note that if we restrict the sample to a given threshold, \eg\ in total mass, we observe that
the average values shown in each plot slightly vary, but the relative differences among the various morphological classes poorly depend on the adopted cut-off level. This suggests that the result is statistically quite robust and not strongly related to the completeness of the data sample. 

To test the statistical robustness of this result we carried out
a bootstrap resampling \citep{Efron1993} aiming at measuring the significance of the difference in the mean mass values for galaxies of morphological type S0 and Sp.  The bootstrap analysis confirms  that the difference between average $M_*$ of S0 and Sp  is highly significant. Only in $\approx$ 1\%\ of 10$^4$ bootstrapped samples   the difference  between   S0 and Sp $M_*$ averages is significant at a confidence level lower than $2\sigma$. The two samples are different also according to a Kolmogorov-Smirov test carried out on each bootstrapped sample: again only $\approx$ 1\%\  of the bootstrapped samples show Kolmogorov-Smirov's $D$\ that implies a significance lower than 2$\sigma$. This result is valid even if a restriction to the  cluster-centric distance  $r < 0.2 R_{200}$ is applied. It is stronger if  $r< 0.6 R_{200}$ because of a larger original sample size. It is also valid if the initial mass of the SSP  at age zero, or the mass   in a nuclear- burning phase of \citep[as defined following ][]{Fritzetal2011} are considered. An additional confirmation is offered by the mass computed from surface photometry $M_\mathrm{ph}$, following  \citet{BelldeJong2001}.  In this case there are significant differences with $M_*$, which for the full WINGS S0 sample follow a law $\delta \log M = \log M_* - \log M_\mathrm{ph} \approx -4.03 + 0.688  \log M_* - (0.0296 \log M_*)^2$, yielding a $\delta \log M \approx -0.25$ at $\log M_* \approx 9$. A similar effect is also seen for E and Sp galaxies, and remains present  if restrictions to $r < 0.2 R_{200}$ and $r< 0.6 R_{200}$ are applied. Nonetheless, a bootstrap applied to $M_\mathrm{ph}$ confirms the same difference and the levels of significance as for $M_*$. 


The differential distribution of the masses for the three morphological types in the different bins of cluster-centric distance  clearly indicate that the higher masses ($>10^{11} M_{\odot}$) are almost exclusively S0's and E's in the $0-0.2$ $D_{BCG}/R_{200}$ bin. In the $0-0.2$ interval of $D_{BCG}/R_{200}$ we count 259 E's, 395 S0's and 135 spirals, so the statistics here is quite robust.
At larger distances the number of galaxies of different morphologies become approximately equal, but the number of E's rapidly decreases.
Then, in the last bin at $0.8-1.0$ $D_{BCG}/R_{200}$, the lack of spatial completeness of the spectroscopic survey starts to be severe, so we cannot trust very much on the observed distribution.

\begin{figure}
\includegraphics[angle=0,scale=0.9]{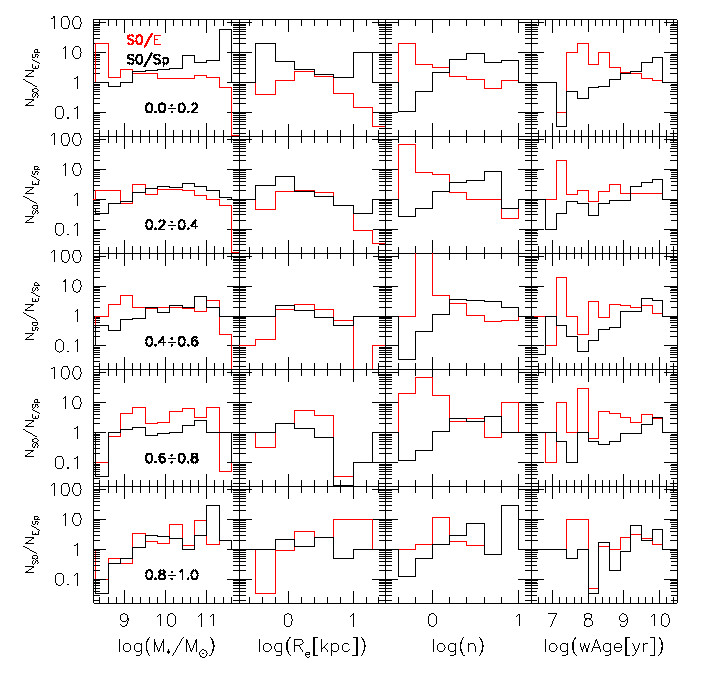}
\caption{From left to right: the log of the total galaxy mass 
($\log(M_{\odot}$), the circularized effective radius ($\log(R^c_e[kpc]$), the log Sersic index ($\log(n)$), and the log of the luminosity weighted age 
($\log(wAge)$). The y-axis shows the ratios between the number of S0's and Spirals (black line) and the number
of S0's and E's (red line) in five different ranges of distances $D_{BCG}/R_{200}$ (0-0.2, 0.2-0-4, 0.4-0.6, 0.6-0.8, and 0.8-1.0).}
\label{fig8}
\end{figure}

\begin{figure}
\includegraphics[angle=0,scale=0.9]{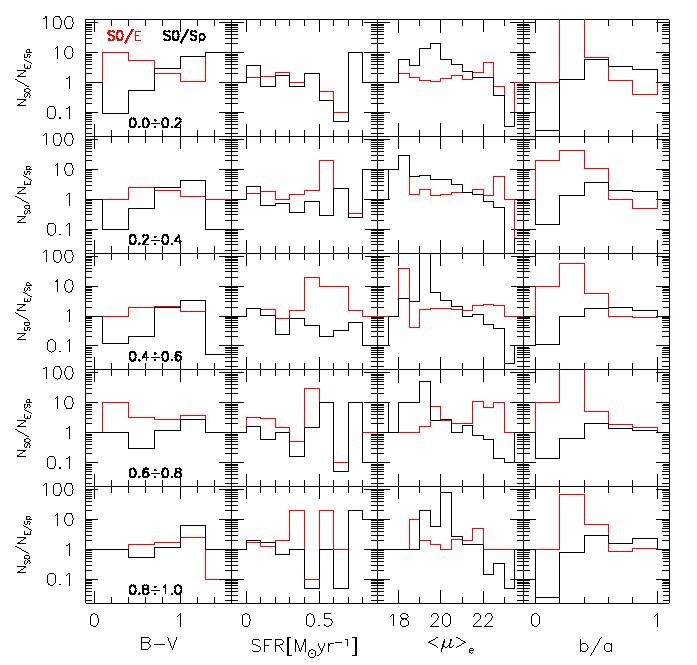}
\caption{From left to right: the B$-$V color, the star formation rate (SFR), the mean effective surface brightness ($\langle \mu \rangle_e$), and the $b/a$ ratio. As in the previous figure the y-axis gives the ratio between the number of S0's and Spirals (black line), and S0's and E's (red line).}
\label{fig9}
\end{figure}

Fig.~\ref{fig8} and Fig.~\ref{fig9} show how the fractions $N_{S0}/N_{Sp}$ and $N_{S0}/N_{E}$ vary as a function of different variables in various
bins of cluster-centric distance. We see
that there are $\sim 50$ times more S0's than Spirals with mass $>10^{11} M_{\odot}$ and up to $\sim 20$ times more S0's than E's of low masses in the first bin. The ratio $N_{S0}/N_{E}$ increases at low masses, but this result is more uncertain being our sample less complete below $\log M_*/M_{\odot} \leq 9.5$. 
In Fig.~\ref{fig8} the distribution of the effective radii $R_e$ looks bimodal for the S0s/Sp ratio, indicating that many of these galaxies are massive, but likely of small dimension, \ie\ quite dense objects, as revealed by their higher surface brightness. They are probably bulge-dominated objects.
It is interesting to note the distributions of the S\'ersic index, of the luminosity weighted age, of the B$-$V color, and of the mean effective surface brightness. All these structural and photometric parameters indicate that in the $0-0.2$ interval of $D_{BCG}/R_{200}$ there is a well defined population
of S0's with $n$ values larger than those of Spirals and lower than those of E's, preferentially older than Spirals, but younger than E's, redder than Spirals but bluer than E's, with a higher surface brightness with respect to Spirals, but similar to E's in all the bins. We also observe that the S0's with small radii have also low $n$ values, young ages, blue colors and SFRs similar to Spirals.
Going at larger cluster-centric distances we instead see a clear trend toward $N_{S0}/N_{Sp}\sim1$ and $N_{S0}/N_{E}\sim1$
in almost all diagrams with the obvious exception of the surface brightness and the axis ratios. The general impression
is that all the galaxies closer to the cluster center are systematically different from the others. This fit with the idea
of a progressive build up of the cluster, where the most processed objects (red and dead) are the first entered in the cluster core, while those in the outer parts are still in a phase of rapid transformation showing traces of residual activity.

\begin{figure}
\includegraphics[angle=0,scale=0.9]{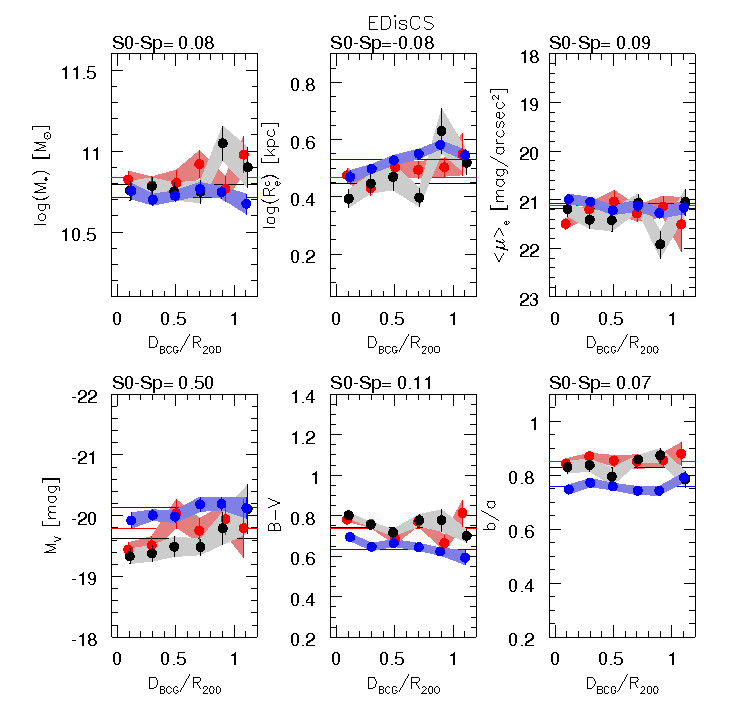}
\caption{The trend of the measured parameters for the EDisCS sample as a function of the cluster-centric distance.}
\label{fig10}
\end{figure}

\begin{figure}
\includegraphics[angle=0,scale=0.9]{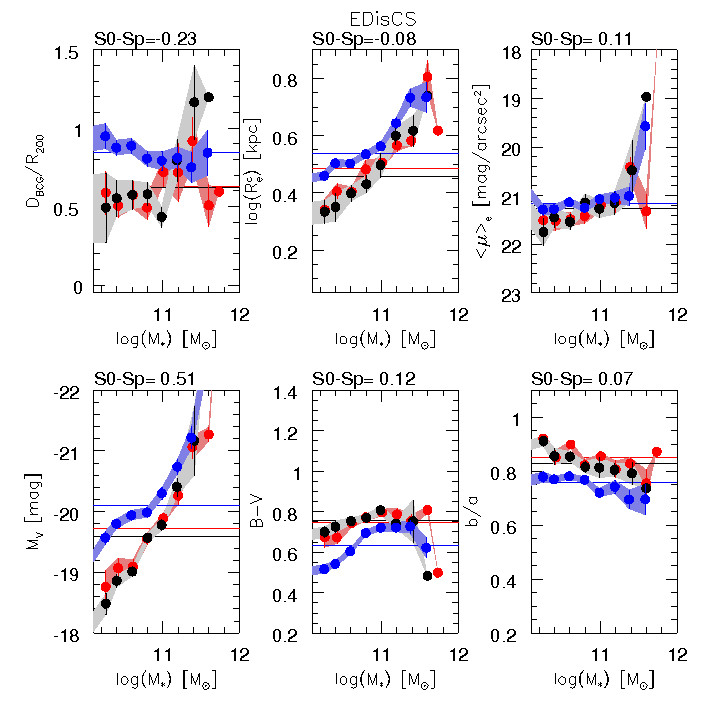}
\caption{The same trends as a function of galaxy masses.}
\label{fig11}
\end{figure}

Looking at the above figures  we must conclude that the effects of galaxy transformation should determine the following changes in the past Spirals (now S0's) with respect to present day Spirals: 1) the mean age, mass and color, 2) the concentration of the light profile so that smaller effective radii and larger surface brightness are measured, 3) the shape of the light profile so that the S\'ersic index increases, 4) progressive cut-off of the SFR. We will discuss these possibilities in Sec.~\ref{Sec6}.

Fig.~\ref{fig10} and Fig.~\ref{fig11} present the behavior of the EDisCS galaxies in various bins of cluster-centric distance and galaxy mass as we did in Fig.~\ref{fig6} and Fig.~\ref{fig7} respectively for the WINGS data. These plots have been obtained in the same way by stacking the EDisCS data as we did for the WINGS sample. Here again the underlying hypothesis is that the properties of the clusters at high redshifts are quite homogeneous and the effects of galaxy evolution not too heavy to change systematically the properties of galaxies in the redshift interval considered. Unfortunately the EDisCS data have not the same degree of accuracy and quality, therefore they are used here only for a qualitative comparison with WINGS.

With respect to WINGS (limited to the same interval of galaxy masses) we note the following behavior at each cluster-centric distance : 1) the masses of S0's and Spirals are nearly equivalent (the mean difference being lower than 0.1 dex for both samples); 2) Spirals are still systematically larger than S0's
in both samples ($S0-Sp=-0.08$ in EDisCS and $S0-Sp=-0.14$ in WINGS). This result however is biased by the fact that the EDisCS effective radii are not deconvolved for the seeing point spread function; 3) in WINGS there is a mean systematic difference in the mean surface brightness of S0's and Spirals of $\sim 0.58$, while it is only $\sim 0.1$ in EDisCS. Again this result is a direct consequence of the measured effective radii; 4) Spirals are systematically brighter than S0's in EDisCS of $\sim 0.5$ mag, while the difference decreases to $\sim 0.2$ in WINGS; 5) the difference in the mean $B-V$ color is nearly equivalent ($\sim 0.1$ mag), while the EDisCS sample is systematically bluer as expected on the basis of their redshift, 6) there is a substantial difference in the mean flattening of Spirals and S0's that we believe could be connected to the difficulty of detecting and classifying nearly edge-on objects at high redshift.

Notably in each bin of galaxy mass the cluster-centric distance of Spirals with respect to S0's and E's is much higher ($S0-Sp=-0.23$ in EDisCS and $S0-Sp=-0.07$ in WINGS). Spirals are also systematically brighter than S0's up to $\sim 10^{11} M_{\odot}$ in EDisCS ($S0-Sp=0.51$ vs $0.16$ in WINGS).

The general impression from these figures is that even at high redshifts the population of S0 galaxies is already quite different from that of Spirals. Once the effects of passive evolution are taken into account, we can say that the Spirals and S0's at high redshift are quite similar to those observed today, apart from their relative frequency.
This means that, either it exists a population of primordial S0's, or the transformation of Spirals
into S0's was active since the first epoch of cluster formation. Furthermore, if a transformation has occurred it should have been quite rapid, probably not longer than $\sim 1-3$ Gyr, because we are not able to detect a class of objects with mixed properties between Spirals and S0's in both surveys. See discussion in Sec.~\ref{Sec6}.

\section{Observational evidences supporting the gas stripping by a hot medium}\label{Sec4}

Ram pressure stripping is one of the physical mechanisms invoked to explain the fast transformation
of Spirals into red passive galaxies similar to many S0's.
Direct evidences that this phenomenon is currently active in the WINGS clusters have been presented by \cite{Poggiantietal15}. They have identified 241 possible Jellyfish galaxies that exhibit material outside the main body of the galaxies, suggesting a phenomenon of gas stripping.

\begin{figure}
\includegraphics[angle=0,scale=0.8]{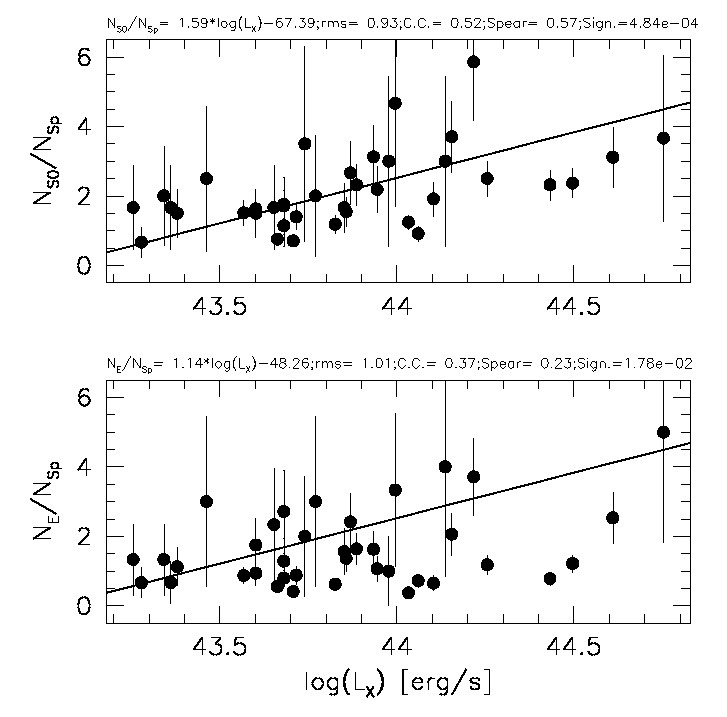}
\caption{The panel shows the X-ray luminosity of the WINGS clusters plotted against the morphological fractions S0/Sp (upper panel) and E/Sp (lower panel). Poissonian error bars have been assumed.}
\label{fig12}
\end{figure}

Another observational fact suggesting that Spirals are affected by hot IGM interaction comes from Fig.~\ref{fig12} that shows the relationship between the X-ray luminosities (in log units) of the WINGS clusters and their morphological fractions S0/Sp and E/Sp.
Here we have taken into account only those WINGS clusters with a robust number of detected morphological types (38 clusters). We have excluded those clusters with less robust statistics in which the number of Spirals members is $\leq 2$. 
The figure clearly indicates that a significant degree of correlation is observed in particular for the S0/Sp fraction. The most luminous X-ray clusters have preferentially a larger S0/Sp (and possibly E/Sp) fraction. This is a clear hint that in the hotter IGM the stripping of gas and consequently the morphological transformation of Spirals is more efficient. The gas is stripped by the hot IGM when Spirals enter the cluster environment and start orbiting around the cluster center. 

The dependence of the morphological fractions on the X-ray luminosity of the clusters, and consequently on the temperature of the intra-cluster medium,  led us to suspect that
the mechanism of evaporation \citep[introduced by][]{Cowie&Songaila1977} might play a role as significant as that of ram pressure. 

\begin{figure}
\center
\includegraphics[angle=0,scale=0.7]{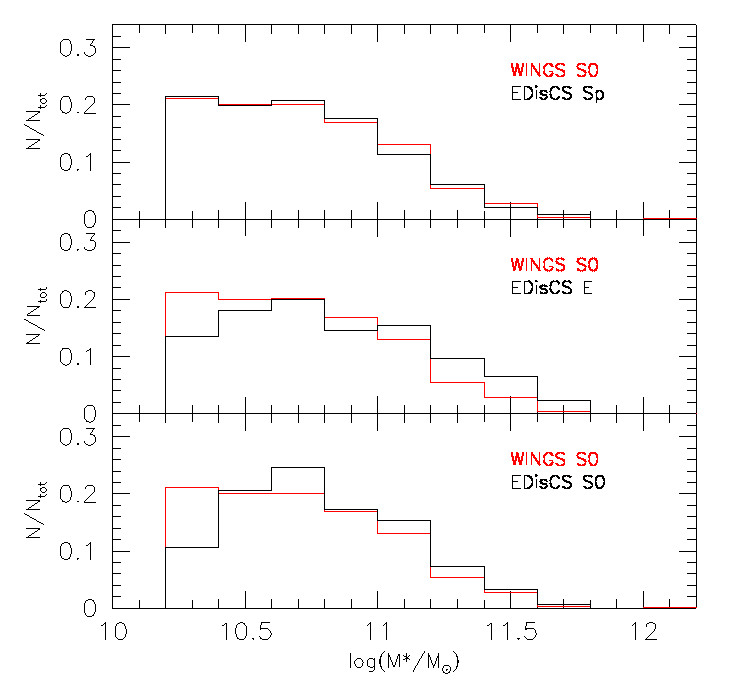}
\caption{From top to bottom the mass distribution of the WINGS S0 galaxies compared with that of Spirals, E's and S0's from the EDisCS galaxies.}
\label{fig13}
\end{figure}

In general, if Spirals transform into S0's via gas stripping and quenching of SF, one might suspect that the mass distribution of present day S0's
should not be too different from the mass distribution of early Spirals. Loosing their gas Spirals do not
change very much their total mass ($<10$\% variations are expected). This is indeed observed in
Fig.~\ref{fig13}.  This figure plots the mass distribution for the two samples of WINGS and EDisCS for galaxy stellar masses above $10^{10.2} M_{\odot}$. Note the strong similarity between the WINGS S0's and EDisCS Spiral population. This similarity is also very clear in Fig. 7 of \cite{Vulcanietal2011} showing the mass function of the two morphological classes in WINGS and EDisCS.
Certainly this is not a conclusive argument, but it contributes to the general picture in favor of the galaxy transformation.

\begin{figure}
\center
\includegraphics[angle=0,scale=0.7]{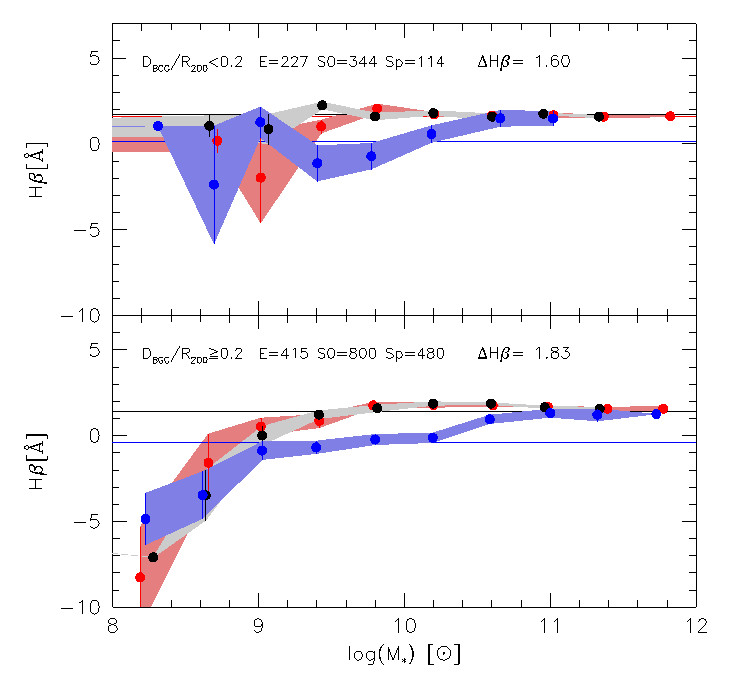}
\caption{The Lick index H$_{\beta}$ for galaxies closer to the center of the clusters and for more distant objects.}
\label{fig14}
\end{figure}

A further evidence that galaxies are stripped of their gas, particularly in the central region of the clusters, comes
from Fig.~\ref{fig14} that shows the behavior of the Lick index H$_{\beta}$ for the WINGS galaxies measured
by Alexander Hannson in his PhD thesis. The figure clearly indicate that the galaxies in the outer region of the clusters
still present this line in emission, while the galaxies in the vicinity of the center are systematically deprived of their gas. 
A more detailed analysis of the spectral emission properties of the WINGS galaxies will be presented by Marziani et al. (in preparation).

Are these hints sufficient to claim that the gas stripping is the only mechanism producing the transformation of Spirals into S0's? Certainly not. In the next section we will see that do exist several suspects
that merging events and gravitational interactions are active somewhere in the clusters and can contribute to the formation of S0 galaxies.

\section{Observational evidences supporting the merging hypothesis}\label{Sec5}

As mentioned in the Introduction, the possibility that S0 galaxies are produced via the merging of two spirals with unequal mass has been discussed since the late 1990s. \cite{Bekki1998} found that  simulations of the merging of two bulgeless galaxies (with the gas component modeled) and  a mass ratio 0.3 could reproduce  the photometric and structural properties of a typical S0 galaxy (red color,  absence of a stellar thin disk).  \cite{Aguerrietal2001} discussed a similar scenario for S0 formation, and found a steepening of the photometric profiles toward the early-types  due to satellite accretion. After a minor merger, bulges grow and the S\'ersic index increases. \cite{Eliche-Moral2013} found that intermediate mergers (i.e., events with mass ratios between 1:4 and 1:7) can produce remnants that photometrically and kinematically resemble S0 galaxies, consistent with the bulge-to-disk coupling observed in real S0's. A significant disk heating is expected from minor mergers \citep{Quetal2011}. 

\begin{figure}
\center
\includegraphics[angle=0,scale=0.7]{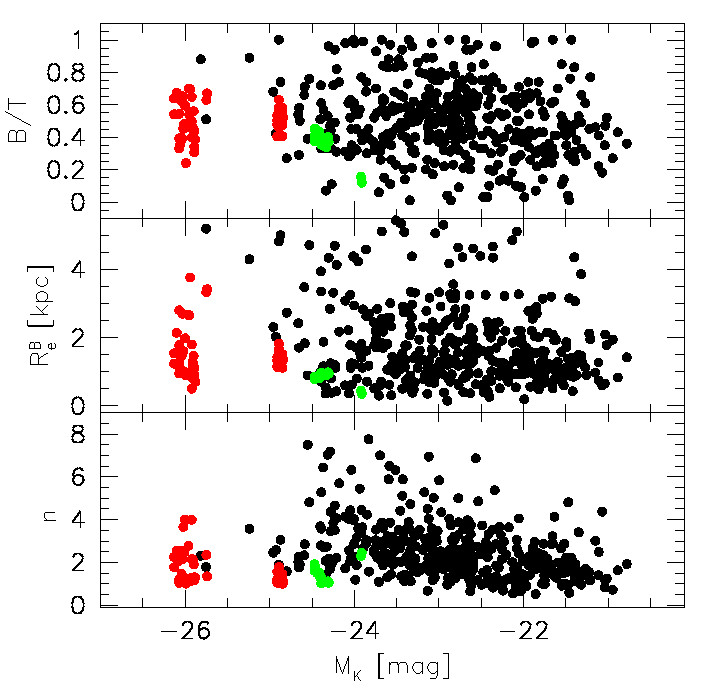}
\caption{The bulge-to-total luminosity ratio B/T, the effective radius of the bulge, and the S\'ersic index of the WINGS S0 galaxies are plotted here versus the K band absolute magnitude. The red and green dots mark the bulge-to-disc decomposition of the merging products coming from N-body simulations of \cite{Querejetaetal2015} and \cite{Tapiaetal2014} respectively.}
\label{fig15}
\end{figure}

These models provide support to the idea that minor/medium mergers lead to an increase of the S\'ersic index value, if spheroids or B+D systems are involved.  The simulations of \cite{Hilzetal13} show  that unequal mergers ($m_{2}< 1/5$) produces a steep increase in the S\'ersic index in massive spheroid ($\log M>11$). Repeated minor merging can even lead to the build-up of elliptical galaxies  without resorting to major mergers \citep{bournaudetal05}. 
\cite{Querejetaetal2015} computed photometric profiles from the  dissipative N-body binary merger simulations from the GalMer database, and found that minor merger remnants can have low S\'ersic indices  $1 < n < 2$, consistent with the presence of pseudo-bulges in observed S0s. 
In general therefore N-body simulations predict that the S\'ersic index of the galaxies evolve, proportionally increasing
with the dimension of the new formed bulge component.

\begin{figure}
\center
\includegraphics[angle=0,scale=0.6]{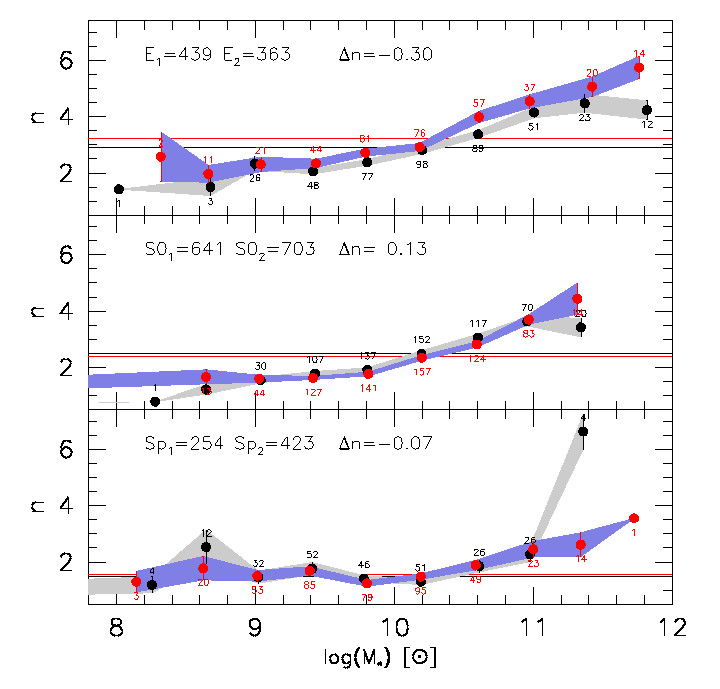}
\caption{The S\'ersic index as a function of galaxy stellar mass for the three morphological types in two different
regimes of cluster-centric distances. Black dots mark the values for
$D_{BCG}/R_{200}\leq 0.3$ while red dots those for $D_{BCG}/R_{200}> 0.3$.}
\label{fig16}
\end{figure}

From an observational point of view we see evidences of ongoing tidal interactions in the
WINGS galaxies \citep[see e.g.][]{Poggiantietal15}. This means that gravitational interactions and merging can still be active today in particular in the outskirts of clusters. Unfortunately the analysis of present day S0's does not easily
reveal whether they are the end product  of past merging events. 
Direct proof of past accretion events could come from the analysis of the frequency of GCs or PN, or from a detailed analysis of the stellar populations and SFH. This is quite difficult already for very nearby objects, but almost impossible for far S0's that can be studied only in their integrated properties. 

The only possible approach with our data is that of statistically
distinguish the mean properties of the different morphological classes, and
to check whether the observed differences might be explained by theoretical models. 
Fig.~\ref{fig15} compares the B/T ratio and the S\'ersic index derived from N-body simulations of major and minor merging events among binary galaxies with the corresponding values measured for the WINGS data set by Sanchez-Jannsen in his PhD thesis. The figure suggests that the formation of cluster S0 galaxies by merging could be one element to consider, at least regarding its brightest representatives. However, it is not clear yet whether smaller cluster S0s can be the end product of merging events.

The problem is that the high velocity dispersion in clusters does not favor the occurrence of merging.
We therefore suspect that most of the merging events happen when galaxies reside in the outskirts of clusters
or when they are still members of small groups that are infalling into the clusters. This means that a substantial
pre-processing could be experienced by Spirals before being virialized into the cluster environment.

\begin{figure}
\center
\includegraphics[angle=0,scale=0.4]{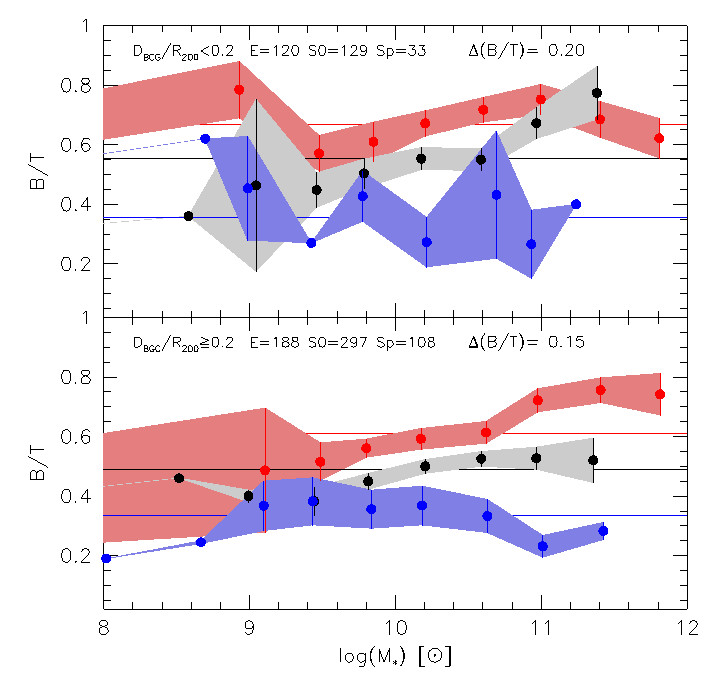}
\includegraphics[angle=0,scale=0.4]{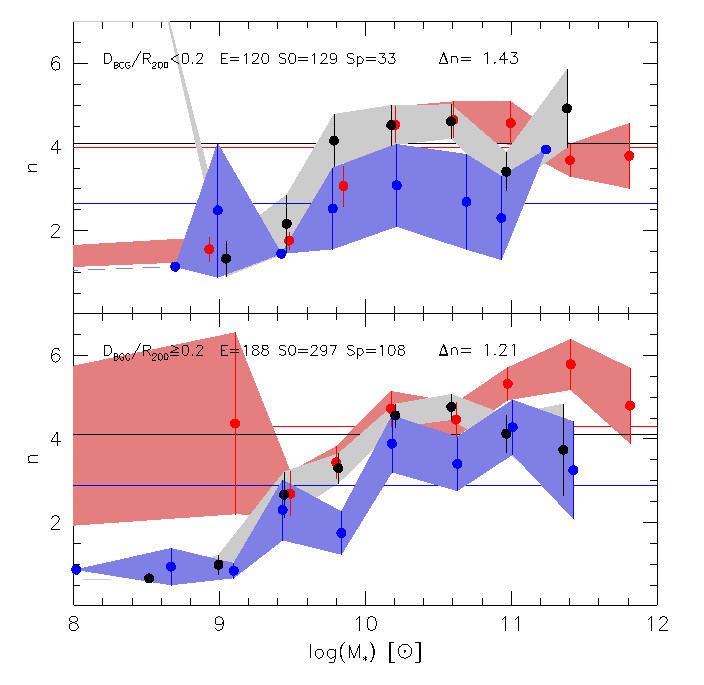}
\caption{{\it Left panel:} The B/T ratios of the various morphological types in the two bin of cluster-centric distance.
{\it Right panel:} The S\'ersic index of the bulge components of the galaxies.}
\label{fig17}
\end{figure}

A second hint in favor of the merging hypothesis comes from Fig.~\ref{fig3}, where we have seen that the fraction S0/E
at low redshift cannot be explained without a small increase of the E types, and this might be explained with major/minor merging events. 

Therefore, prompted by the idea that merging events are necessary for explaining the
actual structure of S0 galaxies and the frequency of the morphological fractions in nearby clusters,
we decided to check if the WINGS data in general, and the S\'ersic index of S0's and Spirals in particular,
are consistent with the merging hypothesis. If such hypothesis is correct,
we expect to see an evolution of the S\'ersic index, i.e. that $n$ increases as a consequence of multiple merging events experienced by Spirals.

In Fig.~\ref{fig6} and  \ref{fig7} we have already shown that the S\'ersic index\footnote{The S\'ersic index was measured by GASPHOT \citep[see,][]{Donofrioetal2014} on the major and minor growth curves of the galaxies without distinguishing the bulge and disk components.} of the various morphological classes are quite distinct both for various masses and for distances from the cluster center. 
The two figures showed that for a relatively broad mass range ($9.5 < log(M_*/M_{\odot}) < 11$) the mean S\'ersic index of S0's and today Spirals has a systematic difference of $\Delta\log(n) \sim 0.2$. In other words, in the same mass range S0's and Spirals have different light profiles. We can distinguish three broad mass domains: small stellar mass $\log(M_*/M_{\odot}) \leq 9.5$: the systematic difference between S0's and Spirals is small. In this mass domain, we have the least massive systems with the smallest bulges, and $n$ is consistent with the presence of pseudo bulges, i.e. of structures that might originate from the secular evolution of disks. In the intermediate mass range, we observe instead systematically higher $n$ for S0's and E's with respect to Spirals. 
For masses $\log(M_*/M_{\odot}) \gtrsim 11$ the $n$ value for Spirals becomes not reliable because there are too few objects of this mass. The general trend is that as far as the spheroidal component of the galaxies increases, the value of $n$ increases.

Trusting in N-body simulations we expect that the larger S\'ersic index of S0's and E's at each mass could be originated through a series of major/minor merging events experienced by a Spiral galaxy. Ram pressure and evaporation cannot in fact drastically change the S\'ersic index of a galaxy. A redistribution or an accretion of new stars is required to produce the observed variations, in particular of the bulge-to-disk ratios. The transforming galaxies should therefore experience such merging events or should produce new stars in their bulges.

\cite{Vulcanietal2011b} came to a similar conclusion in their analysis of the mass functions and morphological mix of the WINGS and EDisCS data.
They stressed that it is impossible to reconcile the mass function observed at low redshift only with the evolution of the morphological fractions.
Mass accretion events and new infalling galaxies are required to reproduce the observed evolution. They believe that new SF events should occur in the transforming galaxies.

A robust hint that merging events and gravitational interactions are responsible of the observed $n-M_*$ relation
would be the detection of a systematic change of $n$ in galaxies that potentially might be subject to several encounters.
We have seen before that the center of clusters, where the density of galaxies and the velocity dispersion is high, host
the more massive objects (a factor of $\sim 2$ larger in mass) as well as the smaller in size ($\sim 25\%$ smaller, with the notable exception of the BCGs). This difference cannot be explained in the context
of ram pressure stripping and evaporation by a hot IGM, but fits with the idea that a series of merging events and strong gravitational interactions have produced the big S0's that are now seen in the cluster center.

In order to test such hypothesis, we examined the behavior of the S\'ersic index of the three morphological classes in Fig.~\ref{fig16}. Here for each class we plotted the mean galaxy mass and S\'erisc index  observed in two
intervals of cluster-centric distance $D_{BCG}/R_{200}\leq 0.3$ (black dots and gray strips) and
$D_{BCG}/R_{200}>0.3$ (red dots and blue strips). The figure shows that there is only a marginal evidence of an increase of $n$ for the S0's in the cluster central region. The probability that the observed means are different by chance
is $\sim 10\%$ and the observed variation is very small. On the other hand the figure shows that inner and outer E's have significantly different values of $n$. In particular the inner E's have systematically lower $n$ than outer objects. The statistical significance that the two samples have different averages by chance is now $\sim 1\%$.
This means that E's in the central region of the clusters are likely subject to tidal stripping (galaxy harassment), i.e. the most external stars are lost producing less extended envelopes and consequently lower measured $n$. Spirals on the other hand remain the same at every distance inside the clusters.

One might speculate that the variations of $n$ could not be associated with a real change of the bulge structure since GASPHOT is a software that provides the
S\'ersic index for the whole galaxy and not for the bulge component.
For this reason we used again the subsample of galaxies of Sanchez-Jannsen, which contains the bulge-to-disk
decomposition of our galaxies. Fig.~\ref{fig17} confirms that: 1)  there is a systematic variation between the S\'ersic index of the bulges of the various morphological types; 2) the mean difference between the B/T ratios of S0's and Spirals
increases in the inner region of the clusters; 3) the mean difference in the S\'ersic index also increases.  

In conclusion we can say that we have only a marginal indication that the S0's bulges are systematically bigger in the central region of the clusters as a consequence of major/minor merging events, but in general the data do not support the idea that merging events are more  frequent in the inner region of clusters with respect to the outskirts. The question is therefore when and where the transforming Spirals acquire the new stars that complete the transition to the S0 class systematically increasing their bulge structures. 

Up to now several works have already claimed that the main
quenching and morphological transformation do not occur primarily in the cores of clusters, at least not at $z < 1$. The decrease in SF is already visible at several virial radii \citep{Lewisetal2002,Gomezetal2003}, and the S0 fraction increases since $z\sim 0.5$ \citep{Dressleretal1997,Fasanoetal2000} particularly in less-massive clusters \citep{Poggiantietal2009,Justetal2010}. The evolution driven by the environment seems to occur at intermediate densities and outside the cluster virial radius and such effects are observed in simulations out to as many as five virial radii \cite{Baheetal2013}. This open the possibility that pre-processing mechanisms
affect the infalling galaxy population. 

The emerging scenario from the above discussion is that Spirals start to quench their SF when they enter in the hot IGM, but at the same time when they are still in the cluster outskirts they progressively accrete/merge new material changing the structure of their bulge and disk (thick disks?). One possibility is that the Spirals which end up in today S0's, were members of small groups infalling the clusters and subsequently merged together. 

In other words from the present sample we can only derive indirect hints that merging events have contributed to the formation of today S0's. 
This il likely due to the fact that the present data cover only the inner virialized part of the WINGS clusters.
Probably the ongoing observations of the outskirts of the WINGS clusters will provide soon much solid conclusions about this.

\begin{figure}
\center
\includegraphics[angle=0,scale=0.40]{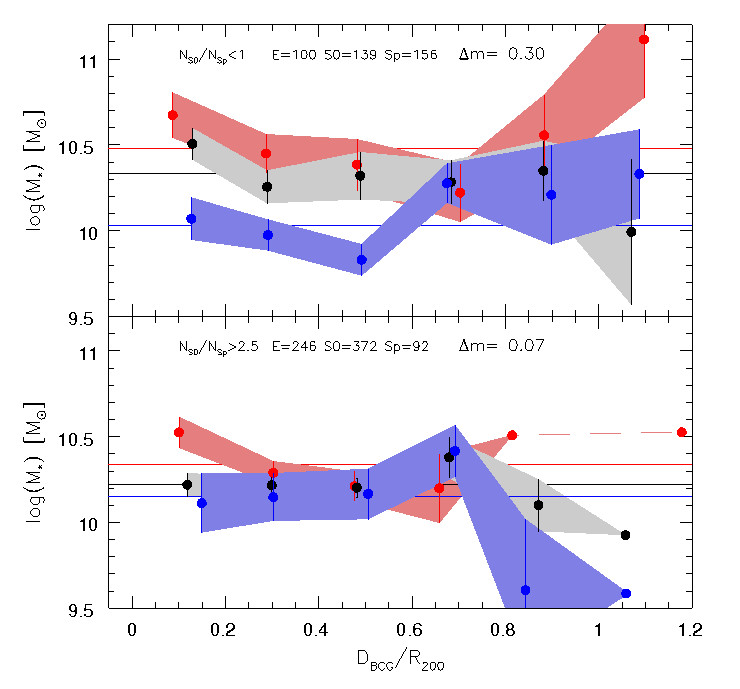}
\includegraphics[angle=0,scale=0.40]{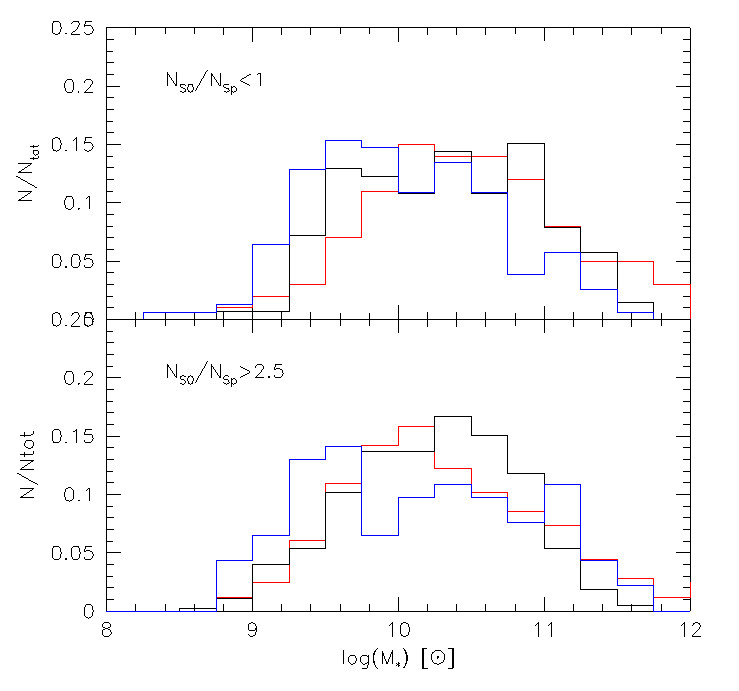}
\caption{{\it Left panel:} The masses of the three morphological types in clusters with a high and small ratio of S0's to Spirals. {\it Right panel:} The distribution of the masses of the three morphological types in clusters with a different fraction of S0/Sp.}
\label{fig18}
\end{figure}

An intriguing fact that is difficult to explain with the ram pressure is that the clusters with a small value of the $N_{S0}/N_{Sp}$ ratio ($<1$) seem to contain in their central region systematically more massive S0's than clusters with a higher value of this ratio (see left panel of Fig.~\ref{fig18}). We have verified that there is not a big difference in the mean luminosity weighted age between these two types of clusters ($\sim 0.1$ dex), so we cannot attribute the effect to a systematic variation in age. 

\begin{figure}
\includegraphics[angle=0,scale=0.8]{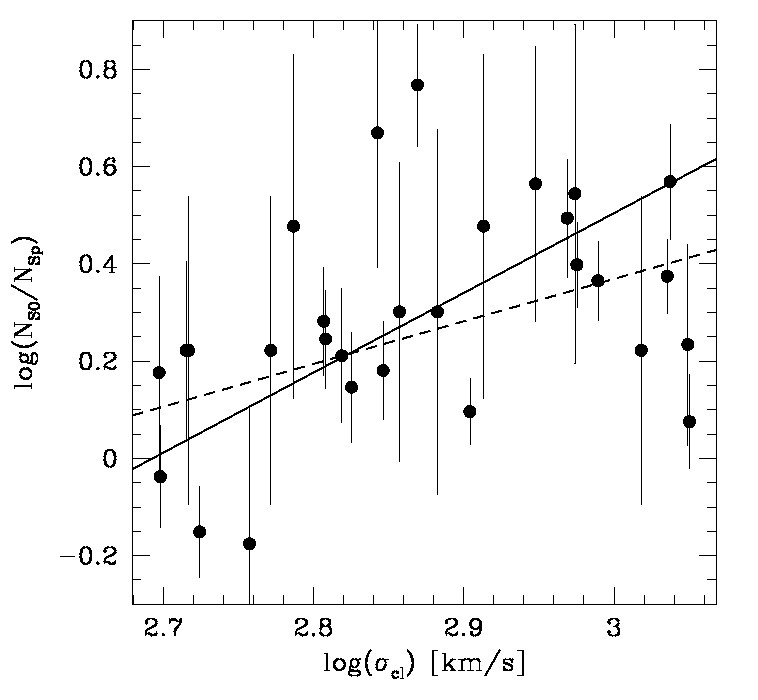}
\caption{The panel show the log of the $N_{S0}/N_{Sp}$ fraction against the central velocity dispersion of the clusters.
The solid line is the bisector least square fit, while the dashed line is the weighted least square fit.}
\label{fig19}
\end{figure}

However, Fig.~\ref{fig19} suggests a partially significant correlation between the log of the $N_{S0}/N_{Sp}$ ratio and the central velocity dispersion of the clusters. A bisector least square fit gives: $\log(N_{S0}/N_{Sp})=1.64\log(\sigma_{cl})-4.42$ with a $rms=0.2$ $c.c.=0.42$ and a significance of $1.79\,10^{-2}$). 
A weighted least square fit provides a shallower slope: $\log(N_{S0}/N_{Sp})=0.87\log(\sigma_{cl})-2.26$. 
On average the fraction $N_{S0}/N_{Sp}$ increase by a factor $2-3$ going from $\sigma_{cl}\sim500$ to $\sigma_{cl}\sim1100$.
This implies that the more massive galaxies reside in the clusters with the lowest
velocity dispersion, in agreement with the idea that merging events are more frequent because the velocity dispersion in the clusters is low.

\cite{Fasanoetal2000} noted that the $N_{S0}/N_{Sp}$ ratio is lower in more concentrated clusters where the number of E's is higher. The suspect is then that interactions and merging have played a key role in determining the high mass of these galaxies. Probably in these clusters the merging activity was higher and the $N_{S0}/N_{Sp}$ ratio kept small 
either by the lower temperature of the X-ray gas (remember that the $N_{S0}/N_{Sp}$ fraction depends also on $L_X$)
and by a continuous accretion of new Spirals from the cosmic web.

The right panel of the same figure shows that there are no systematic differences in the mass distributions of the various morphological types in clusters with $N_{S0}/N_{Sp}<1$ and $N_{S0}/N_{Sp}>2.5$ suggesting that the observed behavior is not the result of a selection bias, but is probably an intrinsic characteristic of galaxies of these two types of clusters.

At the end of this Section we should say that we have accumulated several hints, although none strictly compelling,
that merging events and gravitational interactions (harassment?) have had a role in galaxy clusters (in particular in those
with low velocity dispersions). 

\section{Field and cluster S0's}\label{Sec6}

One of the most relevant things of the previous analysis is that
E's in the central region are systematically deprived of their outer stars with respect to more distant objects. 
This tell us that gravitational stripping and galaxy harassment are playing an active role in the high density regions of clusters.
Galaxy harassment may be associated with the origin of a diffuse stellar component  that yields the so-called intra-cluster optical light (\cite{Zwicky1952}), if this diffuse component is due to stars tidally stripped from galaxies as they move within the cluster \cite[e.g.,][]{GreggandWest1998}.

To test this effect we compare now the properties of the three morphological classes for objects belonging to the general field and to clusters. 
We have therefore used the PM2GC data sample described in Sec.~\ref{sec1c}. 

\begin{figure}
\center
\includegraphics[angle=0,scale=0.40]{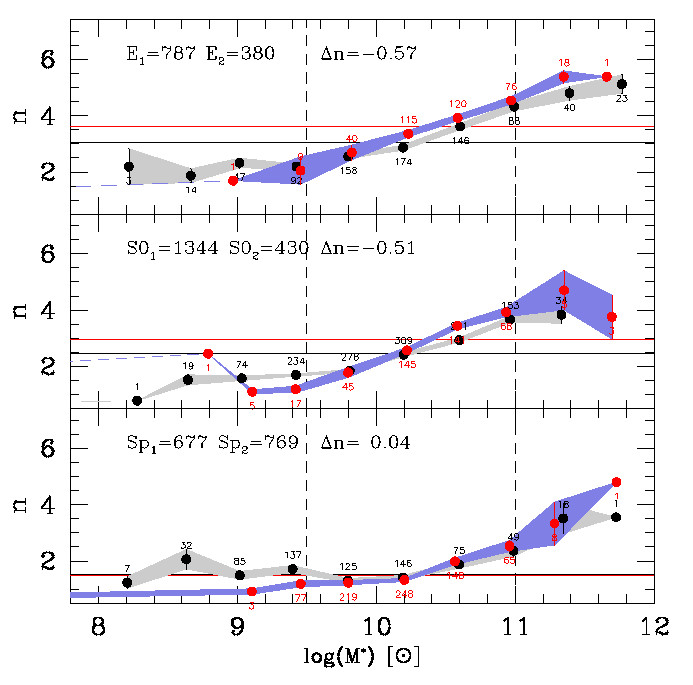}
\includegraphics[angle=0,scale=0.40]{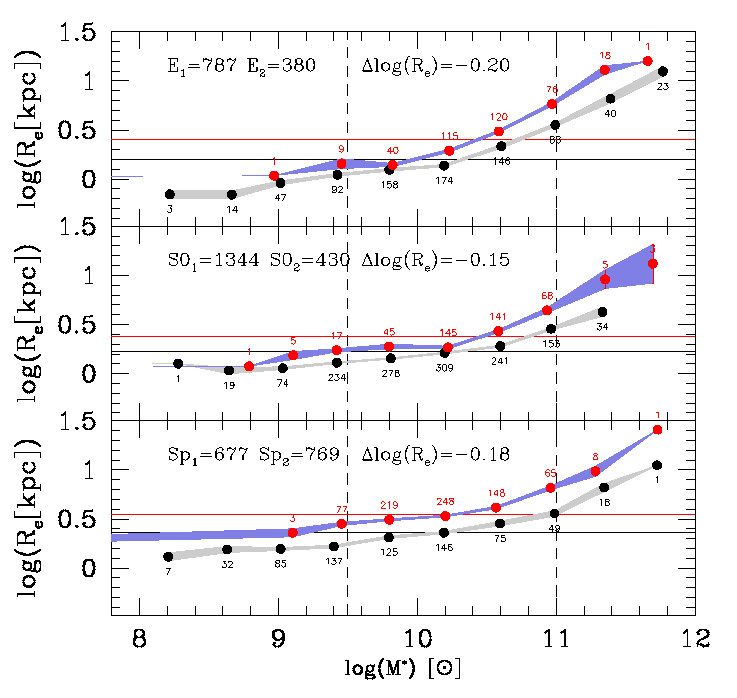}
\caption{{\it Left panel:} The S\'ersic index of E's (upper panel), S0's (middle panel) and Spirals (lower panel) of
galaxies in clusters (back dots and gray strips) and in the field (red dots and blue strips). {\it Right panel:} The effective
radius in log units. Symbols as in the left panel}
\label{fig20}
\end{figure}

Fig.~\ref{fig20} shows the comparison of the S\'ersic index
and the effective radius of the three morphological types for galaxies in clusters (black dots and gray strips) and in the
field (red dots and blue strips). The left panel tells us that Spirals in clusters and in the field have exactly the same
$n$. On the other hand, S0's and E's in clusters appear systematically with lower $n$ values at each mass. We have checked with a Welch's test that this difference is significative from a statistical point of view and cannot originate from the different wavebands (B and V) of the two datasets. The mean difference that can be attributed to the waveband effect
is of the order of $\sim -0.05$ for $\Delta\log(R_e[V-B])$ and $\sim -0.2$ for $\Delta n[V-B]$.
The right panel confirms that the higher S\'ersic index of galaxies in the field is associated to a systematic variation of the circularized effective radius. Field galaxies are systematically larger than cluster objects, suggesting that the main visible action of the cluster environment is the stripping of the outer regions of galaxies.

The main indication is that the mean E's and S0's of intermediate mass in the field have larger $n$ ($\sim 10\%$) and larger $R_e$ ($\sim 20\%$). 
One possibility is that during the infall into the cluster potential, or as a consequence of repeated high speed encounters, galaxies loose their outer regions. This produces systematically smaller radii and also smaller S\'ersic indexes for the galaxies in clusters. Moreover, the similarity of the $n-M_*$ relation of field and cluster galaxies suggests that the mechanisms that have produced the various morphological classes should be very similar in both environments.

If ram pressure and evaporation were not active in the field, the only possible hypotheses are that field S0's originated from Spirals that were members of small groups that are now dissolved, or in binary associations.  Part of these objects could also be byproduct of a secular evolution of the disk. In other words the merging hypothesis is still one the most attractive even for field S0's.

\section{Conclusions and discussion}\label{Sec7}

In this work we have obtained the following results:

\begin{enumerate}
\item the fractions S0/Sp and E/Sp derived for clusters at different redshifts smoothly increase from $z\sim 0.8$ to the present, while the fraction S0/E remains almost constant (or is only slightly increasing) although with a significative variance. In particular the fraction S0/Sp is nearly double every Gyr;
\item starting from the mean number of morphological types observed in EDisCS and ending to those measured in WINGS, we calculated a mean rate of transformation of Spirals into S0's of $\sim 5$ objects per Gyr (not considering the infall of new Spiral objects in the clusters), while the mean transformation of Spirals into E's is at least $\sim 2$ objects per Gyr;
\item the S0 population in clusters appears to grow at the expenses of the Spiral one by a factor of $\sim 3-4$ since $z\sim1$. At the same time the number of E's could not remain constant; 
\item S0's are more frequent in the central region of nearby clusters, while Spirals are preferentially found in the outer regions
of high redshift clusters; 
\item the S0 family at $z\sim0$ is on average quite different from that of present day Spirals: in general at a given distance from the cluster center they are more massive (up to a factor of 2), $25\%$ smaller in size, with an higher S\'ersic index and effective surface brightness, and are redder in B$-$V colors. On the other hand, at a given mass they are closer to the center,
systematically smaller in size, with higher surface brightness, age, color and S\'ersic index, but lower SFR;
\item the more massive S0's in WINGS are clustered around the center of the clusters;
\item the S0's in EDisCS have the same structural characteristics of present day S0's once a passive luminosity evolution is taken into account. As for the WINGS clusters E's and S0's are more frequent in the central part of the clusters, but the proportion of the morphological types is quite different;
\item the mass distribution of present day S0's and high redshift Spirals is very similar (although in the limited range of galaxy masses $M_* > 10^{10.2} M_{\odot}$); 
\item the morphological fractions S0/Sp and E/Sp in the local WINGS sample are mildly correlated with the X-ray luminosity of the clusters and with the cluster velocity dispersion. The objects closer to the cluster center are systematically deprived of their gas. The more massive S0's reside in the clusters with the smaller velocity dispersion;
\item the S\'ersic index of Spirals, S0's and E's increase with the mass of the galaxies and in particular with the bulge component. At each mass the value of $n$ is large for E's, intermediate for S0's and small for Spirals;
\item Pairwise merger simulations give some support to the idea that the most massive cluster S0s may arise by merging;
\item there are no clear indications of an increase of the S\'ersic index for galaxies closer to the cluster center with respect to more distant objects, with the possible exception of central E's (BCGs excluded) that appear to have systematically lower $n$ of far E's;
\item the comparison between cluster and field galaxies tells us that galaxies in clusters are systematically smaller in effective radius and have lower $n$ values. 
\end{enumerate}

The first four points of this list strongly suggest that Spirals transform into S0's and E's in clusters. 

The fifth and seventh points confirm that S0's and Spirals are very different in many observable parameters at all redshifts.
If we accept the idea that Spirals transform into S0's, we should recognize that
the morphological transformation is radical. The speed of the transformation should be quite fast, since the various surveys at different redshifts have clearly identified the three morphological classes with almost no objects with confused or mixed morphology.

The eighth point suggests that the mass distribution of present day S0's and Spiral is consistent (at least for the large masses) with the hypothesis of transformation through gas stripping and quenching of SF. However, within this scenario it is quite difficult to account for the large observed difference in the S\'ersic index among the progenitor Spiral and the final S0 galaxy.

The ninth point tell us that the hot IGM certainly play a key role in the transformation because the clusters having a bright X-ray luminosity contain proportionally more E's and S0's with respect to Spirals. However, the same point tell us that
the clusters with the smaller velocity dispersion contains the biggest S0's in agreement with the idea that merger events
have produced these galaxies.

The activity of merging and gravitational interactions is suggested by points 3, 10 and 13. The main hints that S0's could originate from merging events (if we accept the idea that Spirals must transform), rest on: a) the fact that up to
$D_{BCG}/R_{200}\leq 0.7$ at each distance form the cluster center the mean Spirals are systematically less massive than the mean S0's and E's; b) in the same interval the mean E's are more massive than the mean S0's and the most massive are progressively segregated in the central region; c) the S\'ersic index is predicted to evolve after merging and is observed to increase systematically with the mass of the spheroidal component (and marginally decrease with the distance from the cluster center).

The twelfth point indicates that the merging activity is not confined to the densest regions of the clusters, but can occur in every place and time inside them, in particular in the outer cluster regions, where the velocity dispersion slow down. We can therefore speculate that when small groups fell into the cluster potential they merged together forming the new S0's and E's we see today.  

The thirteenth point suggests that the main action of the cluster environment is that of stripping stars from the outer parts of the galaxies. All the morphological types in clusters appear in fact to be systematically smaller in size and with lower S\'ersic index with respect to field objects. This effect of harassment is not in contradiction with the idea that merging is the main actor of the Spiral transformation.

The merging hypothesis is currently supported by
recent observations of massive S0's located in groups and clusters, usually exhibiting traces of past merger events \citep{Rudiketal2009,Janowieckietal2010,Mihosetal2013}. 
Furthermore, the high resolution analysis of the Atlas3D group \citep{Cappellarietal2011,Cappellarietal2013,Emsellemetal2011} has found that 87\% of the galaxies classified as S0's or E's at very low redshifts are fast rotating systems, indicating that they have rotationally supported disks. Slow rotators are the most massive galaxies, often with boxy isophotes (70\% of them), and depletion of mass in the central galaxy regions (core). Fast rotators are less massive, typically have disky isophotes, and centrally peaked surface brightness profiles (cuspy).  Fast rotators are presumably stripped Spirals with some mass added by minor mergers, whereas the slow rotators are formed in more violent processes so that the bulges formed first.

This scenario is also consistent with other well known observations:  1) stellar and gaseous disks, and dust lanes have been frequently found in E's and S0' bulges; 2) shells and ripples are often seen around these galaxies when deep exposures are taken; 3) lenticular galaxies surrounded by polar rings exist; 4) early-type galaxies could evolve in size by multiple minor merging events \citep[see \eg,][]{Lopezetal2012}.  

All these facts together point toward a strong role of merging in re-structuring the morphology of late-type galaxies.
This is supported by modern high resolution simulations of merging events \citep[e.g.][]{Coxetal2006,Robertsonetal2006} between two Spiral galaxies that have produced E's and S0's with some rotational support with the right coupled bulge-disk structures in time scale less than $\sim$ 3 Gyr \citep{Querejetaetal2015}. A major merging could completely transform the structure of a Spiral galaxy, varying its bulge-to-disk ratio, its S\'ersic index, the mean surface brightness and the effective radius \cite[see \eg,][]{Querejetaetal2015}.
Minor mergers can also produce S0's with intermediate kinematic properties
between fast and slow rotators that are difficult to explain with major mergers \citep{Tapiaetal2014}. In conclusion the bulges of S0's can be formed by means of a major merger of early disks or by a sequence of minor merger events, followed by a later disk rebuilding of the left-over gas and stripped stars
\citep{SomervillePrimack1999}.

The data presented here are consistent with this scenario, although there are only hints that this was the right story for the S0 in clusters.
Direct evidences exist that ram pressure stripping and thermal evaporation could rapidly quench the SF in Spiral galaxies, while internal secular evolution can re-distribute the
matter within them.



\section*{Acknowledgments}
We thank Benedetta Vulcani for providing us the masses of the EDisCS galaxies in advance of publication. We also thank Alexander Hannson and Ruben Sanchez-Janssen for the permission of using the data of their PhD theses.

\bibliographystyle{frontiersinSCNS_ENG_HUMS} 









\end{document}